\documentclass[pra,twocolumn,showpacs,amsmath,amssymb,superscriptaddress]{revtex4-1}

\usepackage{graphicx, amsmath, bm, braket, txfonts, color, bbold}

\begin{document}

\title{Acoustic phonons and strain in core/shell nanowires}

\author{Christoph Kloeffel}
\affiliation{Department of Physics, University of Basel, Klingelbergstrasse 82, CH-4056 Basel, Switzerland}
\author{Mircea Trif}
\affiliation{Department of Physics and Astronomy, University of California, Los Angeles, California 90095, USA}
\affiliation{Laboratoire de Physique des Solides, CNRS, Universit\'{e} Paris-Sud, 91405 Orsay, France}
\author{Daniel Loss}
\affiliation{Department of Physics, University of Basel, Klingelbergstrasse 82, CH-4056 Basel, Switzerland}

\date{\today}

\begin{abstract}

We study theoretically the low-energy phonons and the static strain in cylindrical core/shell nanowires (NWs). Assuming pseudomorphic growth, isotropic media, and a force-free wire surface, we derive algebraic expressions for the dispersion relations, the displacement fields, and the stress and strain components from linear elasticity theory. Our results apply to NWs with arbitrary radii and arbitrary elastic constants for both core and shell. The expressions for the static strain are consistent with experiments, simulations, and previous analytical investigations; those for phonons are consistent with known results for homogeneous NWs. Among other things, we show that the dispersion relations of the torsional, longitudinal, and flexural modes change differently with the relative shell thickness, and we identify new terms in the corresponding strain tensors that are absent for uncapped NWs. We illustrate our results via the example of Ge/Si core/shell NWs and demonstrate that shell-induced strain has large effects on the hole spectrum of these systems. 

\end{abstract}

\pacs{63.22.Gh, 63.22.-m, 62.20.-x, 71.70.Fk}

\maketitle

\section{Introduction}
\label{sec:Introduction}

In the past years, it has been demonstrated that the performance of nanowires (NWs) can greatly benefit from the presence of a shell. For instance, surface passivation is an option to reduce scattering, and measurements on InAs/InP core/shell NWs have revealed significantly higher mobilities than in uncapped InAs NWs \cite{vantilburg:sst10}. Furthermore, experiments on various core/shell NWs \cite{treu:nlett13, mayer:nco13, hocevar:apl13, skoeld:nlett05, rigutti:prb11} have demonstrated that adding a shell can be very useful for optical applications, a feature that is well-known, e.g., from colloidal quantum dots (QDs) \cite{mahler:nma08, chen:nma13}. In Ge/Si core/shell NWs, which recently attracted attention, the shell is beneficial for several reasons. In particular, it provides a large valence band offset at the interface, leading to a strongly confined hole gas inside the Ge core without the need for dopants \cite{lu:pnas05, park:nlt10}. 

Due to lattice mismatch, core/shell heterostructures are usually strained, which can have important consequences on their electrical and optical properties. For instance, strain may affect the lifetimes of spin qubits \cite{maier:prb12, maier:prb13} and has already been used to tune photons from separate QDs into resonance \cite{flagg:prl10, kuklewicz:nlett12}. The reason for such effects lies in the strain dependence of the Hamiltonian of the electronic states \cite{birpikus:book}. For the core/shell NWs of Refs.~\onlinecite{treu:nlett13, skoeld:nlett05, hocevar:apl13, rigutti:prb11}, a strong and strain-based correlation between the shell thickness (or composition) and the wavelength of the emitted photons has already been measured. In addition, the shell-induced strain may lift quasidegeneracies in the spectrum of NWs and NW QDs \cite{kloeffel:prb11}. Considering these and other possibly relevant consequences, knowledge of the strain distribution in core/shell NWs is crucial. Exact calculations of the lattice displacement, however, typically require numerics \cite{groenqvist:jap09, hestroffer:nanotech10, hocevar:apl13}. Analytical results are rare and require simplifying assumptions that may or may not be justified, depending on the choice of materials and on the effects that one is interested in. The model of Ref.~\onlinecite{hestroffer:nanotech10}, for instance, assumes purely uniaxial strain along the NW axis. The results of Refs.~\onlinecite{liang:jap05, schmidt:prb08} apply when the core and shell materials are isotropic and have the same elastic properties. To our knowledge, the most general formulas provided so far are those of Ref.~\onlinecite{menendez:anphbe11}, assuming isotropic media and requiring only Poisson's ratio to be the same in core and shell.

A particularly attractive feature of NWs is their potential to host electrically controllable spin qubits \cite{loss:pra98}. While bare InAs and InSb NWs have been the workhorse systems \cite{fasth:prl07, nadjperge:nat10, schroer:prl11, vandenberg:prl13, kloeffel:annurev13}, spin qubits may also be implemented in core/shell NWs such as Ge/Si \cite{hu:nna07, roddaro:prl08, hu:nna12, higginbotham:arX14}, for which a large degree of external control has been predicted \cite{kloeffel:prb13}. As electrons and holes interact with lattice vibrations, understanding of the quantum mechanical behavior of the system requires knowledge of the phonon bath. For instance, it is well-known that phonons can be dominant decay channels for spin qubits \cite{khaetskii:prb01, golovach:prl04, kroutvar:nat04, hanson:rmp07, hachiya:arX, kornich:prb14}. The shell of core/shell NWs not only induces static strain, it also affects the phonon modes. While phonons in homogeneous NWs have been discussed in detail in the literature \cite{cleland:book, landau:elasticity}, we are not aware of analytical results for NWs with a finite shell.   

In this paper, we derive algebraic expressions for the static strain and the low-energetic phonon modes in core/shell NWs. Assuming isotropic materials and a force-free wire surface, we allow for arbitrary core and shell radii, independent elastic properties in core and shell, and take all components of the stress and strain tensors into account. Our results for the phonons illustrate that the shell notably affects the phonon-based displacement fields and, among other things, that the dispersion relations of the longitudinal, torsional, and flexural modes change differently with the shell thickness. In particular, new terms arise in the corresponding strain and stress tensors that are absent in homogeneous NWs. We illustrate our results via the example of Ge/Si NWs, given the fact that the coherence of their interfaces has already been demonstrated experimentally \cite{dillen:prb12, dillen:nna14}. The derived formulas for the static strain can be considered a further extension of those listed in Ref.~\onlinecite{menendez:anphbe11} and are consistent with experiments \cite{dillen:prb12, dillen:nna14, rigutti:prb11, hestroffer:nanotech10, treu:nlett13, skoeld:nlett05, hocevar:apl13} and numerical simulations \cite{groenqvist:jap09, hestroffer:nanotech10, hocevar:apl13, rigutti:prb11}. We calculate the effects of the static strain on the low-energy hole spectrum of Ge/Si NWs, complementing the analysis of Ref.~\onlinecite{kloeffel:prb11}. 

The paper is organized as follows. In Sec.~\ref{sec:LinearElasticityTheory} we introduce the notation and recall relevant relations from linear elasticity theory. The results for the static strain are derived in Sec.~\ref{sec:StaticStrainMain}, where we also investigate the effects of strain on the spectrum of Ge/Si NWs. Having summarized the low-energetic phonon modes in homogeneous NWs in Sec.~\ref{sec:PhononsBareWireMain}, we extend these results to the case of core/shell NWs in Sec.~\ref{sec:PhononsCoreShellWireMain}, followed by concluding remarks in Sec.~\ref{sec:ConclusionsOutlook}. The appendix contains useful relations and further details of the calculations. In particular, providing also a comparison to the case of bulk material, we discuss the displacement operator and the normalization condition for phonons in core/shell and core/multishell NWs, as quantization is mandatory for quantum mechanical analyses.

\section{Linear elasticity theory}
\label{sec:LinearElasticityTheory}

In this section we recall relevant relations from linear elasticity theory and introduce the notation used throughout this paper. The information summarized here is carefully explained in Refs.~\onlinecite{cleland:book, landau:elasticity}, and we refer to these for further details. 

In a bulk semiconductor without additional forces, the atoms form a periodic and very well structured lattice, characterized by the lattice constant $a$. The displacement of an atom $X$ from its original position $\bm{r}_X$ is described by the displacement vector $\bm{u}(\bm{r}_X)$. It may be caused by externally applied forces, or, as in the case of core/shell NWs, by an interface between materials with different lattice constants. In the continuum model, the displacement field $\bm{u}(\bm{r})$ is directly related to the strain tensor elements $\epsilon_{ij}$ via 
\begin{equation}
\epsilon_{ij}=\frac{1}{2}\left(\frac{\partial u_i}{\partial x_j}+\frac{\partial u_j}{\partial x_i}\right),
\end{equation}
leading to strain-induced effects on the conduction and valence band states \cite{birpikus:book}. The position is denoted here by $\bm{r} = \sum_i x_i \bm{e}_i$, where the three orthonormal basis vectors $\bm{e}_i$ are the unit vectors along the orthogonal axes $i$. Important quantities besides the strain are the stress tensor elements $\sigma_{ij}$. These are of relevance as $\sigma_{ij}dA$ corresponds to the force along $\bm{e}_i$ experienced by an area $dA$ normal to $\bm{e}_j$. We note that $\epsilon_{ij} = \epsilon_{ji}$ and $\sigma_{ij} = \sigma_{ji}$, which implies that the strain and stress tensors are fully described by three diagonal and three off-diagonal elements each.  

For semiconductors with diamond (Ge, Si, \ldots) or zinc blende (GaAs, InAs, \ldots) structure, the relation between stress and strain is given by
\begin{equation}
\begin{pmatrix}
\sigma_{11} \\ \sigma_{22} \\ \sigma_{33} \\ \sigma_{23} \\ \sigma_{13} \\ \sigma_{12}
\end{pmatrix} = 
\begin{pmatrix}
c_{11} & c_{12} & c_{12} & 0 & 0 & 0 \\ 
c_{12} & c_{11} & c_{12} & 0 & 0 & 0 \\
c_{12} & c_{12} & c_{11} & 0 & 0 & 0 \\
0 & 0 & 0 & c_{44} & 0 & 0 \\
0 & 0 & 0 & 0 & c_{44} & 0 \\
0 & 0 & 0 & 0 & 0 & c_{44}
\end{pmatrix}
\begin{pmatrix}
\epsilon_{11} \\ \epsilon_{22} \\ \epsilon_{33} \\ 2 \epsilon_{23} \\ 2 \epsilon_{13} \\ 2 \epsilon_{12}
\end{pmatrix} ,
\label{eq:StressStrainRelDiamondZB}
\end{equation} 
where the $c_{ij}$ are the elastic stiffness coefficients and $\{1,2,3\}$ are the main crystallographic axes. Calculations with the exact stiffness matrix often require elaborate numerical analyses, and it is therefore common to replace the stress-strain relations by those of an isotropic material. This simplification usually results in good approximations when compared with the precise simulations \cite{cleland:book, landau:elasticity}. The elastic properties of such a material are fully described by the two Lam\'{e} parameters $\lambda$ and $\mu$, which are found from Young's modulus $Y$ (often denoted by $E$) and Poisson's ratio $\nu$ through
\begin{gather}
\lambda = \frac{Y\nu}{(1+\nu)(1-2\nu)}, \label{eq:LambdaYoungsModPoissonsRat} \\
\mu = \frac{Y}{2(1+\nu)}. \label{eq:MuYoungsModPoissonsRat}
\end{gather} 
Considering the limit of isotropic media, we thus replace the stiffness coefficients of Eq.\ (\ref{eq:StressStrainRelDiamondZB}) by $c_{12} = \lambda$, $c_{44} = \mu$, and $c_{11} = 2\mu + \lambda$. The relations between stress and strain are now the same for arbitrarily rotated coordinate systems. Hence, referring to NWs, we obtain 
\begin{equation}
\begin{pmatrix}
\sigma_{rr} \\ \sigma_{\phi\phi} \\ \sigma_{zz} \\ \sigma_{\phi z} \\ \sigma_{rz} \\ \sigma_{r\phi}
\end{pmatrix} = 
\begin{pmatrix}
2\mu + \lambda & \lambda & \lambda & 0 & 0 & 0 \\ 
\lambda & 2\mu + \lambda & \lambda & 0 & 0 & 0 \\
\lambda & \lambda & 2\mu + \lambda & 0 & 0 & 0 \\
0 & 0 & 0 & \mu & 0 & 0 \\
0 & 0 & 0 & 0 & \mu & 0 \\
0 & 0 & 0 & 0 & 0 & \mu
\end{pmatrix}
\begin{pmatrix}
\epsilon_{rr} \\ \epsilon_{\phi\phi} \\ \epsilon_{zz} \\ 2 \epsilon_{\phi z} \\ 2 \epsilon_{rz} \\ 2 \epsilon_{r\phi}
\end{pmatrix} ,
\label{eq:StressStrainRelNW}
\end{equation}       
where 
\begin{eqnarray}
\bm{e}_r &=& \bm{e}_x \cos{\phi} + \bm{e}_y \sin{\phi} , \\
\bm{e}_\phi &=& - \bm{e}_x \sin{\phi} + \bm{e}_y \cos{\phi} ,
\end{eqnarray}
and $\bm{e}_z = \bm{e}_x \times \bm{e}_y = \bm{e}_r \times \bm{e}_\phi$ are the orthonormal basis vectors for cylindrical coordinates $r$, $\phi$, and $z$. The vector $\bm{e}_z$ is oriented along the symmetry axis of the NW, while $\bm{e}_r$ and $\bm{e}_\phi$ point in the radial and azimuthal direction, respectively. From 
\begin{equation}
\bm{r} = r \bm{e}_r + z \bm{e}_z = x \bm{e}_x + y \bm{e}_y + z \bm{e}_z
\end{equation}
it is evident that the cartesian coordinates $x$ and $y$ are related to $r$ and $\phi$ through $x = r \cos{\phi}$ and $y = r \sin{\phi}$ (the $z$ axis is the same in both coordinate systems). We wish to emphasize that we use
\begin{equation}
r = \sqrt{x^2 + y^2}
\end{equation}
throughout this work in order to avoid confusion with the density $\rho$, and so $r \neq |\bm{r}|$. Detailed information about the stress and strain tensor elements in cartesian and cylindrical coordinates is provided in Appendixes \ref{secsub:Transformation2ndRankTensors} and \ref{secsub:StrainInCylindricalCoords}. Finally, we note that Eq.~(\ref{eq:StressStrainRelNW}) is independent of the growth direction of the NW because of the isotropic approximation.

\section{Static strain in core/shell nanowires}
\label{sec:StaticStrainMain} 

An interface between two materials of mismatched lattice constants induces a static strain field. In core/shell NWs, such an interface is present at the core radius $R_c$. When the lattice constants in core ($a_c$) and shell ($a_s$) are different, the system will tend to match these for reasons of energy minimization. For instance, $a_c = 5.66\mbox{ \AA}$ and $a_s = 5.43\mbox{ \AA}$ for Ge/Si core/shell NWs \cite{winkler:book, adachi:properties}, and so the shell tends to compress the core lattice, strongly affecting the properties of the confined hole gas \cite{kloeffel:prb11, birpikus:book}. Below, we analyze the strain in core/shell NWs and derive algebraic expressions for both the inner and outer part of the heterostructure. The resulting static strain is found by assuming a coherent interface between the two materials, i.e., pseudomorphic growth. We consider the limit of an infinite wire, which applies well away from the ends of the NW \cite{groenqvist:jap09} when the length $L$ is much larger than the shell radius $R_s$ ($L \gg R_s$). Our approach is similar to those used previously \cite{liang:jap05, schmidt:prb08, groenqvist:jap09, hestroffer:nanotech10, menendez:anphbe11}.

\subsection{Boundary conditions}

When the strain changes slowly on the scale of the lattice spacing, the displacement field leads to the distorted lattice vectors \cite{groenqvist:jap09}
\begin{equation}
\bm{l}_{c,s}(\bm{r},\bm{n})= \sum_{i} \bm{e}_i \sum_j \left(\delta_{ij}+\frac{\partial u^{c,s}_i(\bm{r})}{\partial x_j}\right)n_j a_{c,s}
\label{eq:distorted_lattice_vecs}
\end{equation}
when viewed from an atom at position $\bm{r} + \bm{u}^{c,s}(\bm{r})$ in the core ($c$) or shell ($s$), respectively, where $\bm{e}_i$ are the orthonormal basis vectors of the lattice, $\bm{n}$ is a vector with integer components $n_i$, and $\delta_{ij}$ is the Kronecker delta. Pseudomorphic strain requires the components of the distorted lattice vectors in core and shell that are parallel to the interface to match at the core-shell transition. Thus, $\bm{t}\cdot\bm{l}_{c} = \bm{t}\cdot\bm{l}_s$, where $\bm{t}$ stands for an arbitrary tangent to the core-shell interface. Using cylindrical coordinates, the two orthogonal directions $\bm{e}_\phi$ and $\bm{e}_z$ are the basis vectors for any $\bm{t}$, which results in the boundary conditions $\bm{e}_\phi \cdot \bm{l}_{c} = \bm{e}_\phi \cdot\bm{l}_s$ and $\bm{e}_z \cdot \bm{l}_{c} = \bm{e}_z \cdot\bm{l}_s$ at radius $R_c$. Furthermore, the core-shell transition needs to be spatially matched in the radial direction, i.e., there should be no unrealistic gaps or overlaps between the two materials at the interface. 

In order to ensure pseudomorphic strain, we start from an initial configuration where the shell is unstrained and the core is highly strained, such that the lattice constant of the core matches the one in the shell \cite{povolotskyi:jap06, groenqvist:jap09}. Of course, this initial arrangement is unstable, and so the system will relax into a stable and energetically favored configuration. Considering the continuum limit, the requirements for a coherent interface at $r \simeq R_c$ can now be summarized in a simple form \cite{nishiguchi:prb94},
\begin{equation}
\tilde{\bm{u}}^c(R_c, \phi, z) = \tilde{\bm{u}}^s(R_c, \phi, z) ,
\label{eq:boundaryCond_contDispl}
\end{equation}
where the $\tilde{\bm{u}}^{c,s}$, in contrast to the $\bm{u}^{c,s}$ of Eq.\ (\ref{eq:distorted_lattice_vecs}), denote the displacement from the initially matched configuration. That is, 
\begin{equation}
\bm{u}^{c}(\bm{r}) = \tilde{\bm{u}}^{c}(a_s \bm{r}/a_c) + \frac{a_s - a_c}{a_c}\bm{r}
\label{eq:ucanductildeMain}
\end{equation}
and $\bm{u}^{s}(\bm{r}) = \tilde{\bm{u}}^{s}(\bm{r})$. As illustrated in Eq.~(\ref{eq:ucanductildeMain}), the displacement field $\bm{u}^{c}(\bm{r})$ in the core can be described by a sum of two parts. When the lattice constant changes from $a_c$ to $a_s$, the term $(a_s/a_c - 1)\bm{r}$ shifts an atom that is originally located at $\bm{r}$ to its new position $a_s\bm{r}/a_c$. Additional displacement from this new position is then accounted for by $\tilde{\bm{u}}^{c}(a_s \bm{r}/a_c)$. Consequently, the strain tensor elements $\epsilon^{c,s}_{ij}(\bm{r})$ in the core and shell read 
\begin{eqnarray}
\epsilon^{c,s}_{ij}(\bm{r}) &=&\frac{a_s}{2 a_{c,s}}\Biggl[ \frac{\partial \tilde{u}_i^{c,s}(\bm{r}^\prime)}{\partial x_j^\prime} + \frac{\partial \tilde{u}_j^{c,s}(\bm{r}^\prime)}{\partial x_i^\prime} \Biggr]_{\mbox{\footnotesize{$\bm{r}^\prime = a_s\bm{r}/a_{c,s}$}}} \nonumber \\ & & + \frac{a_s - a_{c,s}}{a_{c,s}} \delta_{ij},
\label{eq:strainElemsCoreShellMain}
\end{eqnarray}
where $\bm{r}^\prime = \sum_i x_i^\prime \bm{e}_i$. The resulting strain tensor $\epsilon^{c,s}(\bm{r})$ is linearly related to the stress tensor $\sigma^{c,s}(\bm{r})$ via the Lam\'e constants [Eq.\ (\ref{eq:StressStrainRelNW})]. As additional boundary conditions, the stress must be continuous at the interface \cite{nishiguchi:prb94} and we assume that the shell surface is free of forces,
\begin{eqnarray}
\sigma^{c}(R_c, \phi, z) \bm{e}_r &=& \sigma^{s}(R_c, \phi, z) \bm{e}_r ,
\label{eq:boundaryCond_contStress} \\
\sigma^{s}(R_s, \phi, z) \bm{e}_r &=& 0.
\label{eq:boundaryCond_noStress}
\end{eqnarray}
Next, using the above-mentioned boundary conditions, we derive algebraic expressions for the static strain in core/shell NWs.

\subsection{Analytical results}
\label{secsub:StatStrainAnalyticalResults}
 
From symmetry considerations, the displacement in both core and shell must be of the form
\begin{equation}
\tilde{\bm{u}}^{p} = \tilde{u}_r^{p}(r) \bm{e}_r + \eta_z^{p}(r) z \bm{e}_z,   
\end{equation}
where we introduced $p \in \{c,s\}$ for convenience. For the displacement field to be static, the differential equations
\begin{equation}
\sum_j \frac{\partial \sigma_{ij}^p}{\partial x_j} = 0   
\end{equation}
need to be solved in the absence of body forces \cite{landau:elasticity}, and in doing so we find
\begin{gather}
\tilde{u}_r^{p} = \alpha_{p} r + \frac{\beta_{p}}{r} - \frac{\delta_{p} (\lambda_{p} + \mu_{p})}{2(\lambda_{p} + 2 \mu_{p})} r \ln{r}, \\
\eta_z^{p} = \gamma_{p} + \delta_{p} \ln{r}, 
\end{gather} 
where $\lambda_{c,s}$ and $\mu_{c,s}$ are the Lam\'e parameters in the core and shell, respectively. The coefficients $\alpha_{c,s}$ to $\delta_{c,s}$ are to be determined from the boundary conditions. Since the displacement must be finite in the center, we first conclude that $\beta_{c} = 0 = \delta_{c}$. Second, also $\delta_{s} = 0$ because $\sigma_{rz}^{s} \propto \delta_s z/r$ must vanish at the surface [Eq.~(\ref{eq:boundaryCond_noStress})], and so the $\eta_z^{c,s} = \gamma_{c,s}$ are constants. Consequently, one obtains
\begin{equation}
\gamma_s = \epsilon^{s}_{zz} = \frac{a_e - a_s}{a_s},
\end{equation}   
with $a_e$ as the resulting (equilibrium) lattice constant in the $z$ direction. From the boundary conditions listed in Eqs.\ (\ref{eq:boundaryCond_contDispl}), (\ref{eq:boundaryCond_contStress}), and (\ref{eq:boundaryCond_noStress}), we can express all nonzero coefficients 
\begin{gather}
\gamma_c = \gamma_s, \\
\beta_s = (\alpha_c - \alpha_s) R_c^2 ,
\end{gather}
and $\alpha_{c,s}$ in terms of $a_e$ only. The latter can finally be found by minimizing the elastic energy of the system. 

Using Eq.~(\ref{eq:strainElemsCoreShellMain}), the above-mentioned results for $\tilde{\bm{u}}^{c,s}$, and the equations of Appendix \ref{secsub:StrainInCylindricalCoords} for the strain tensor in cylindrical coordinates, one finds
\begin{gather}
\epsilon^{c}_{rr} = \epsilon^{c}_{\phi\phi} = \frac{a_s}{a_c} \alpha_c + \epsilon_0 , \label{eq:ecrrStaticMaintext}\\
\epsilon^{c}_{zz} = \frac{a_s}{a_c} \gamma_c + \epsilon_0 = \frac{a_e - a_c}{a_c}  \label{eq:eczzStaticMaintext}
\end{gather}
in the core, and
\begin{gather}
\epsilon^{s}_{rr} = \alpha_s - \frac{\beta_s}{r^2} , \\
\epsilon^{s}_{\phi\phi} = \alpha_s + \frac{\beta_s}{r^2} , \\
\epsilon^{s}_{zz} = \gamma_s = \frac{a_e - a_s}{a_s} 
\end{gather}
in the shell, with $0 = \epsilon^{c,s}_{r\phi} = \epsilon^{c,s}_{rz} = \epsilon^{c,s}_{\phi z}$. The parameter
\begin{equation}
\epsilon_0 = \frac{a_s - a_c}{a_c}
\label{eq:epsilon0maintext}
\end{equation} 
introduced in Eqs.~(\ref{eq:ecrrStaticMaintext}) and (\ref{eq:eczzStaticMaintext}) is the relative mismatch of the lattice constants. Remarkably, Eqs.~(\ref{eq:ecrrStaticMaintext}) and (\ref{eq:eczzStaticMaintext}) imply that the stress and strain are constant within the core, which is consistent with simulations \cite{groenqvist:jap09, hocevar:apl13, hestroffer:nanotech10}. Furthermore, we note that $\epsilon^{c}_{rr} = \epsilon^{c}_{\phi\phi} = \epsilon^{c}_{xx} = \epsilon^{c}_{yy}$. Below, we outline the calculation of $a_e$ and provide our final results.

The elastic energy density in an isotropic solid is \cite{landau:elasticity}
\begin{equation}
F = \frac{1}{2}\sum_{i,j} \sigma_{ij}\epsilon_{ij} = \frac{\lambda}{2}\left[\mbox{Tr}(\epsilon) \right]^2 + \mu \sum_{i,j} \epsilon_{ij}^2 ,
\end{equation}
where $\mbox{Tr}(\epsilon) = \sum_i \epsilon_{ii}$ is the trace of the strain tensor, and 
\begin{equation}
U = \int_V d^{3}\bm{r} F(\bm{r})
\end{equation}
is the elastic energy of an object with volume $V$. For the static strain discussed in this section, the elastic energy density $F_c$ in the core is constant and $F_s$ in the shell depends solely on the coordinate $r$. The elastic energy of the NW can therefore be calculated via
\begin{equation}
U = 2 \pi L \left( \frac{R_c^2 F_c }{2} + \int_{R_c}^{R_s} dr r F_s(r) \right).  
\end{equation}       
By imposing the condition $\partial U / \partial a_e = 0$ in order to find the energetically favored configuration, we obtain algebraic expressions for $a_e$ and, thus, for all previously discussed quantities. As expected, these are not affected by the length of the wire, since $U \propto L$ for the regime $L \gg R_s$ considered here. Moreover, the coefficients $\alpha_{c,s}$ and $\gamma_{c,s}$ do not depend on the absolute values of $R_c$ and $R_s$. Instead, they depend on the relative shell thickness
\begin{equation}
\gamma = \frac{R_s - R_c}{R_c}.
\label{eq:gammaRelShThDefinition}
\end{equation}  
More precisely, it is possible to write the dependence of $\alpha_{c,s}$ and $\gamma_{c,s}$ on the radii in terms of
\begin{equation}
\widetilde{\gamma} = \gamma^2 + 2 \gamma = \frac{R_s^2 - R_c^2}{R_c^2}
\label{eq:gammaTildeDefinition}
\end{equation}
only, which is the ratio between the shell and core area in the cross section. Similarly, the coefficients $\alpha_{c,s}$, $\beta_s$, and $\gamma_{c,s}$ do not depend on the absolute values of the lattice constants $a_c$ and $a_s$, but on the relative lattice mismatch $\epsilon_0$ [Eq.~(\ref{eq:epsilon0maintext})]. We note that $|\epsilon_0| \ll 1$, which is important for the linear elasticity theory of this work to hold. 

The full results are rather lengthy. Nevertheless, they can be very well approximated through an expansion in the small parameter $\epsilon_0$. Neglecting corrections of order $\epsilon_0^2$ and rewriting the results in a convenient form, we obtain
\begin{gather}
\alpha_s = - \frac{\epsilon_0 \xi_c}{2 D_{\rm str}} \left[ \mu_c (2 \mu_s - \lambda_s) + (2 \mu_s^2 - \lambda_s \mu_c) \widetilde{\gamma} \right] , \\
\beta_s = - \frac{R_c^2 \epsilon_0 \xi_c \xi_s}{2 D_{\rm str}} \left[ \mu_c + (\mu_c + \mu_s) \widetilde{\gamma} + \mu_s \widetilde{\gamma}^2 \right] , \\
\gamma_s = - \frac{\epsilon_0 \xi_c}{D_{\rm str}} \left[ \mu_c (2 \mu_s + \lambda_s) + (\lambda_s \mu_c + \mu_s \mu_c + \mu_s^2) \widetilde{\gamma} \right] ,
\end{gather}       
where we defined
\begin{equation}
\xi_p = 2 \mu_p + 3 \lambda_p .
\end{equation}
The denominator is
\begin{eqnarray}
D_{\rm str} &=& \mu_c \xi_c (2 \mu_s + \lambda_s) + \mu_s \xi_s (\lambda_c + \mu_c + \mu_s) \widetilde{\gamma}^2 \nonumber \\
& & + \left[ \xi_c (\lambda_s \mu_c + \mu_s \mu_c + \mu_s^2) + \xi_s \mu_s (2 \mu_c + \lambda_c) \right] \widetilde{\gamma} .
\end{eqnarray}
Inserting $\alpha_c = \alpha_s + \beta_s/R_c^2$, $\gamma_c = \gamma_s$, and the above-listed expressions into Eqs.~(\ref{eq:ecrrStaticMaintext}) and (\ref{eq:eczzStaticMaintext}), and neglecting again terms of order $\epsilon_0^2$, one finds
\begin{gather}
\epsilon^c_{\perp} = \epsilon^c_{rr} = \epsilon^c_{\phi\phi} = \frac{\epsilon_0 \xi_s \mu_s}{2 D_{\rm str}} \left[ (2 \mu_c - \lambda_c) \widetilde{\gamma} + (2 \mu_s - \lambda_c) \widetilde{\gamma}^2 \right] , \label{eq:ecrrFinalResultMain} \\
\epsilon^c_{zz} = \frac{\epsilon_0 \xi_s \mu_s}{D_{\rm str}} \left[ (2 \mu_c + \lambda_c) \widetilde{\gamma} + (\lambda_c + \mu_c + \mu_s) \widetilde{\gamma}^2 \right]  \label{eq:eczzFinalResultMain}
\end{gather}
for the strain in the core.

The expressions derived in this work are more general than those provided previously \cite{hestroffer:nanotech10, liang:jap05, schmidt:prb08, menendez:anphbe11} and may be interpreted as a further extension of those in Ref.~\onlinecite{menendez:anphbe11}. Indeed, by writing the Lam\'{e} parameters $\lambda_{c,s}$ and $\mu_{c,s}$ in terms of Young's modulus $Y_{c,s}$ and Poisson's ratio $\nu_{c,s}$ for core and shell [Eqs.~(\ref{eq:LambdaYoungsModPoissonsRat}) and (\ref{eq:MuYoungsModPoissonsRat})], we find that our results are exactly identical to those of Ref.~\onlinecite{menendez:anphbe11} for the case $\nu_c = \nu_s$ assumed therein. From the ratio of Eqs.~(\ref{eq:ecrrFinalResultMain}) and (\ref{eq:eczzFinalResultMain}),
\begin{equation}
\frac{\epsilon^c_{\perp}}{\epsilon^c_{zz}} = \frac{2 \mu_c - \lambda_c + (2 \mu_s - \lambda_c) \widetilde{\gamma} }{4 \mu_c + 2 \lambda_c + 2 (\lambda_c + \mu_c + \mu_s) \widetilde{\gamma} } ,
\end{equation}
we find that typically $|\epsilon_{zz}^c| > |\epsilon_{\perp}^c|$, in agreement with numerical results \cite{hocevar:apl13, hestroffer:nanotech10, rigutti:prb11, groenqvist:jap09}. Our formulas also feature the correct limits. For instance, we obtain $a_e \rightarrow a_c$ for $\gamma \to 0$ as expected, and so the strain in the core vanishes for a negligibly thin shell. In the limit of an infinite shell ($\gamma \to \infty$), our formulas yield $a_e \rightarrow a_s$, $\epsilon^{c}_{zz} \to \epsilon_0$, and 
\begin{eqnarray}
\lim_{\gamma \to \infty}  \epsilon^{c}_{\perp} = \frac{\epsilon_0 (2\mu_s - \lambda_c)}{2(\lambda_c + \mu_c + \mu_s)} + \mathcal{O}(\epsilon_0^2),
\end{eqnarray}
which corresponds exactly to the previously studied case of a wire embedded in an infinite matrix \cite{yang:prb97}. When switching to cartesian coordinates, we obtain $\epsilon_{xy}^s = -\beta_s \sin(2\phi)/r^2$, which seems consistent with numerics \cite{groenqvist:jap09}. As pointed out in Ref.~\onlinecite{groenqvist:jap09}, where core/shell NWs with anisotropic materials have been investigated, also other off-diagonal strain tensor components may be nonzero in reality. However, these were found to be very small, particularly in the core.

\subsection{Results for Ge/Si core/shell nanowires}
\label{secsub:StatStrainResultsGeSiNWs}

We conclude our discussion of the static stress and strain fields by applying our results to the example of Ge/Si core/shell NWs, demonstrating that the strain can have major effects on the electronic properties of a system. As illustrated in this example, the strain is usually not negligible, and so our formulas derived for both core and shell may prove very useful for a wide range of material combinations.  

\begin{figure}[t]
\begin{center}
\includegraphics[width=0.85\linewidth]{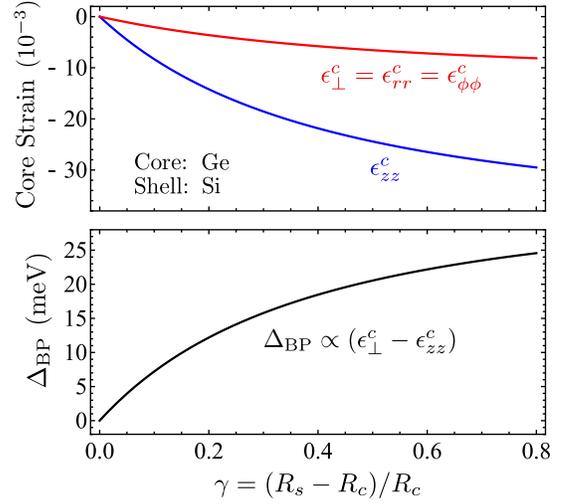}
\caption{Static core strain and its effect on the hole spectrum of Ge/Si core/shell NWs as a function of the relative shell thickness $\gamma$. In the top figure, the nonzero components $\epsilon_{\perp}^c$ and $\epsilon_{zz}^c$ of Eqs.~(\ref{eq:ecrrFinalResultMain}) and (\ref{eq:eczzFinalResultMain}) are plotted for the parameters in the text. The resulting splitting $\Delta_{\rm BP}$ [Eq.~(\ref{eq:DeltaBPFinalMain}), bottom figure] can be as large as \mbox{$\sim$30 meV} and strongly affects the low-energetic hole states, as explained in detail in Ref.~\onlinecite{kloeffel:prb11} [wherein $\delta(\gamma) = \Delta_{\rm BP}(\gamma)$].}
\label{fig:StaticStrainResults}
\end{center}
\end{figure}

Ge/Si core/shell NWs have attracted attention because they host strongly confined hole states inside their cores \cite{lu:pnas05}. Thus, we focus here on the static strain in the core and discuss its effects on the holes in more detail. The lattice mismatch for Ge/Si core/shell NWs is $\epsilon_0 = -0.040$ \cite{winkler:book, adachi:properties}, and the Lam\'{e} constants, listed in units of $10^9\mbox{ N/m$^2$}$, are $\lambda_c = 39.8$, $\mu_c = 55.6$, $\lambda_s = 54.5$, and $\mu_s = 67.5$ (see also Appendix \ref{sec:ParametersGeSiCoreShell}) \cite{cleland:book, adachi:properties}. In Fig.~\ref{fig:StaticStrainResults} (top), the strain tensor elements $\epsilon_{\perp}^c$ and $\epsilon_{zz}^c$ of Eqs.~(\ref{eq:ecrrFinalResultMain}) and (\ref{eq:eczzFinalResultMain}) are plotted as a function of $\gamma$. These are negative over the entire range of shell thicknesses, and so the core material is compressed, as expected from $a_c > a_s$. The dependence of the strain on $\gamma$ is consistent with simulations and experiments \cite{hestroffer:nanotech10}, and we note that $|\epsilon_{zz}^c| > |\epsilon_{\perp}^c|$ for any $\gamma$. 

The effects of strain on hole states in the topmost valence band of Ge are described by the Bir-Pikus Hamiltonian \cite{birpikus:book}. Using the spherical approximation $d = \sqrt{3} b$, which applies well to Ge ($b \simeq -2.5\mbox{ eV}$, $d \simeq -5.0\mbox{ eV}$ \cite{birpikus:book}), and neglecting global shifts in energy, the Bir-Pikus Hamiltonian for holes reads
\begin{equation}
H_{\rm BP} = b \left[\sum \limits_{i} \epsilon_{ii} J_i^2 + 2 \Big( \epsilon_{xy} \{ J_x, J_y \} + \mbox{c.p.}\Big) \right] .
\end{equation} 
Here $b$ and $d$ are the deformation potentials, $J_i$ are the components of the effective spin 3/2 along the axes $i$, \mbox{``c.p.''} stands for cyclic permutations, and $\{A, B\} = (AB + BA)/2$. We note that the axes $x$, $y$, and $z$ need not coincide with the main crystallographic axes due to the spherical approximation. Our results for the static strain in core/shell NWs reveal that the relations $\epsilon^{c}_{xx} = \epsilon^{c}_{yy} = \epsilon^{c}_{\perp}$ and $0 = \epsilon^{c}_{xy} = \epsilon^{c}_{xz} = \epsilon^{c}_{yz}$ are fulfilled in the core. Exploiting these properties and the equality $J_x^2 + J_y^2 = 15/4 - J_z^2$, the Bir-Pikus Hamiltonian for the core is of the simple form
\begin{equation}
H_{\rm BP}^{c} = b \left( \epsilon^{c}_{zz} - \epsilon^{c}_{\perp} \right) J_z^2  ,
\end{equation}   
where global shifts in energy have again been omitted. As discussed in Ref.~\onlinecite{kloeffel:prb11}, this Hamiltonian has important effects on the hole spectrum in Ge/Si core/shell NWs, because it determines the splitting $\Delta$ between the ground states $\ket{g_\pm}$ and the first excited states $\ket{e_\pm}$ at wave number $k_z = 0$ along the NW. The subscripts ``$+$'' and ``$-$'' refer to the spin states, and we note that the total angular momentum along the wire is $F_z = \mp 1/2$ for $\ket{g_\pm}$, whereas $F_z = \pm 1/2$ for $\ket{e_\pm}$. The splitting $\Delta = \Delta_{\rm LK} + \Delta_{\rm BP}$ comprises a strain-independent term $\Delta_{\rm LK} \propto R_c^{-2}$, which arises from the radial confinement and the kinetic energy (Luttinger-Kohn Hamiltonian), and the strain-induced term
\begin{equation}
\Delta_{\rm BP} = \bra{e_\pm} H_{\rm BP}^{c} \ket{e_\pm} - \bra{g_\pm} H_{\rm BP}^{c} \ket{g_\pm} .
\end{equation}
Defining
\begin{equation}
\Lambda = \bra{e_\pm} J_z^2 \ket{e_\pm} - \bra{g_\pm} J_z^2 \ket{g_\pm} ,
\end{equation}
one finally obtains
\begin{equation}
\Delta_{\rm BP} =  b \Lambda \left( \epsilon^{c}_{zz} - \epsilon^{c}_{\perp} \right) .
\label{eq:DeltaBPFinalMain}
\end{equation}
The parameter $\Lambda$ turns out to be independent of $R_c$, and using $\ket{g_\pm}$ and $\ket{e_\pm}$ of Ref.~\onlinecite{kloeffel:prb11} we find $\Lambda = 0.46$ for Ge. That is, $\Delta_{\rm BP} = \Delta_{\rm BP}(\gamma)$ is determined by $\epsilon^{c}_{zz} - \epsilon^{c}_{\perp}$ and depends only on the relative shell thickness.

Figure~\ref{fig:StaticStrainResults} (bottom) shows the dependence of $\Delta_{\rm BP}$ on $\gamma$ for Ge/Si core/shell NWs. Remarkably, $\Delta_{\rm BP}$ can be as large as $\mbox{30 meV}$ and exceeds $10\mbox{ meV}$ at relatively thin shells ($\gamma \gtrsim 0.15$) already. For comparison, one finds $\Delta_{\rm LK} \simeq \mbox{3.0--0.75 meV}$ for typical core radii $R_c \simeq \mbox{5--10 nm}$. Therefore, the splitting $\Delta$ is mostly determined by $\Delta_{\rm BP}$, i.e., by the relative shell thickness. The combination of a small $\Delta_{\rm LK}$ and $0 \leq \Delta_{\rm BP}(\gamma) \lesssim \mbox{30 meV}$ is of great importance not only for the spectrum in the wire, but also for, e.g., the properties of hole-spin qubits in NW QDs \cite{kloeffel:prb11, maier:prb13, kloeffel:annurev13, kloeffel:prb13}.

\section{Phonons in homogeneous nanowires}
\label{sec:PhononsBareWireMain}

In this section, we recall the calculation of lattice vibrations in homogeneous NWs \cite{landau:elasticity, cleland:book, stroscio:book, trif:prb08} and provide the displacement vectors and the core strain for the phonon modes of lowest energy. The information forms a basis for Sec.~\ref{sec:PhononsCoreShellWireMain}, where the analysis is extended to the case of core/shell NWs.

\subsection{Equation of motion, ansatz, and boundary conditions}
\label{secsub:PhononsEOMAnsatzBoundCond}

For an isotropic material with density $\rho$ and Lam\'{e} parameters $\lambda$ and $\mu$, the equations of motion
\begin{equation}
\rho \ddot{u}_i = \sum_j \frac{\partial \sigma_{ij}}{\partial x_j} 
\label{eq:DynEqMostGenMain}  
\end{equation}
can be summarized in the form
\begin{equation}
\rho \ddot{\bm{u}} = (\lambda + \mu) \nabla (\nabla \cdot \bm{u}) + \mu \nabla^2 \bm{u} ,  
\label{eq:DynEqIsotrSolidMain}
\end{equation}
where
\begin{equation}
\nabla = \bm{e}_x \partial_x + \bm{e}_y \partial_y + \bm{e}_z \partial_z = \bm{e}_r \partial_r + \bm{e}_\phi \frac{1}{r} \partial_\phi + \bm{e}_z \partial_z 
\end{equation}
is the Nabla operator and
\begin{equation}
\nabla^2 = \partial_x^2 + \partial_y^2 + \partial_z^2 = \partial_r^2 + \frac{1}{r} \partial_r + \frac{1}{r^2} \partial_\phi^2 + \partial_z^2 
\end{equation}
is the Laplacian. In order to find the eigenmodes for the cylindrical NW, the displacement vector may be written in terms of three scalar functions $\Phi_\eta$, $\eta \in \{0,1,2 \}$, via \cite{stroscio:book, nishiguchi:prb94, sirenko:pre96}
\begin{equation}
\bm{u} = \nabla \Phi_0 + \nabla \times \bm{e}_z \Phi_1 + \nabla \times ( \nabla \times \bm{e}_z \Phi_2) .
\label{eq:AnsatzUthreePotentialsMain}
\end{equation}
Inserting Eq.~(\ref{eq:AnsatzUthreePotentialsMain}) into Eq.~(\ref{eq:DynEqIsotrSolidMain}) and exploiting identities such as $\nabla \cdot (\nabla \times \bm{A}) = 0$, the resulting equation of motion reads
\begin{eqnarray}
0 &=& \nabla \left(\rho \ddot{\Phi}_0 - (2 \mu + \lambda) \nabla^2 \Phi_0 \right) + \nabla \times \bm{e}_z \left(\rho \ddot{\Phi}_1 - \mu \nabla^2 \Phi_1 \right) \nonumber \\
& & + \nabla \times \left[\nabla \times \bm{e}_z \left(\rho \ddot{\Phi}_2 - \mu \nabla^2 \Phi_2 \right)\right] .
\end{eqnarray}
This equation is therefore satisfied when the scalar functions obey the wave equations
\begin{equation}
\rho \ddot{\Phi}_\eta = \left[\mu + \delta_{\eta,0}(\mu + \lambda) \right] \nabla^2 \Phi_\eta ,
\label{eq:WaveEquationsSkalarPotsMain}
\end{equation}
where $\delta_{\eta,0}$ is a Kronecker delta. While these wave equations are sufficient criteria for the equation of motion to be satisfied, we note that some special solutions of Eq.~(\ref{eq:DynEqIsotrSolidMain}) can be found that do not obey Eq.~(\ref{eq:WaveEquationsSkalarPotsMain}). An example is provided below for the torsional mode. However, we also illustrate that this solution can be interpreted as the limit of a more general solution obtained with an ansatz that relies on the above-mentioned wave equations.  

Due to the cylindrical symmetry and the translational invariance along the NW axis $z$ ($L \gg$ radius), the $\Phi_\eta$ can be written in the form
\begin{equation}
\Phi_\eta = f_{\eta}(r) e^{i(q_z z + n \phi - \omega \tau)},
\label{eq:PhiGeneralFormMain}
\end{equation} 
where $q_z$ is the wave number along the wire, $n$ is an integer, $\omega$ is the angular frequency, and $\tau$ is the time. Insertion of Eq.~(\ref{eq:PhiGeneralFormMain}) into Eq.~(\ref{eq:WaveEquationsSkalarPotsMain}) results in the differential equation
\begin{equation}
\left( \partial_r^2 + \frac{1}{r}\partial_r - \frac{n^2}{r^2} - q_z^2 + \frac{\rho \omega^2}{\mu + \delta_{\eta,0}(\mu + \lambda)} \right) f_{\eta}(r) = 0 
\end{equation}  
for the function $f_{\eta}(r)$. With $J_n$ and $Y_n$ as Bessel functions of the first and second kind, respectively, and with $\chi_{\eta, J}$ and $\chi_{\eta,Y}$ as (dimensionful) complex coefficients, the general solution of this differential equation is
\begin{equation}
f_{\eta}(r) = \chi_{\eta, J} J_n (\kappa_{\eta,J} r) + \chi_{\eta, Y} Y_n (\kappa_{\eta,Y} r) ,
\label{eq:fetaofrBasicJYGeneralFormMain}
\end{equation}
where 
\begin{equation}
\kappa_{\eta,J}^2 = \kappa_{\eta,Y}^2 = \frac{\rho \omega^2}{\mu + \delta_{\eta,0}(\mu + \lambda)}  - q_z^2 .
\label{eq:kappaConditionBasicMain}
\end{equation}
We mention that $\kappa_{\eta,J}$ and $\kappa_{\eta,Y}$ need not be identical and may be chosen arbitrarily, provided that Eq.~(\ref{eq:kappaConditionBasicMain}) is satisfied. For homogeneous NWs considered in this section, $\chi_{\eta, Y} = 0$ because $Y_n$ diverges in the limit $r \to 0$, and so
\begin{equation}
\Phi_\eta =  \chi_{\eta, J} J_n (\kappa_{\eta,J} r) e^{i(q_z z + n \phi - \omega \tau)} .
\end{equation}
For given $n$ and $q_z$, the corresponding eigenfrequencies and coefficients can be determined from the boundary conditions that we discuss below.

In vector notation, with the three components referring to $\bm{e}_r$, $\bm{e}_\phi$, and $\bm{e}_z$, respectively, Eq.~(\ref{eq:AnsatzUthreePotentialsMain}) reads as
\begin{equation}
\bm{u} = \begin{pmatrix} u_r \\ u_\phi \\ u_z \end{pmatrix} = \begin{pmatrix} \partial_r \Phi_0 + \frac{1}{r} \partial_\phi \Phi_1 + \partial_r \partial_z \Phi_2 \\ 
\frac{1}{r} \partial_\phi \Phi_0 - \partial_r \Phi_1 + \frac{1}{r} \partial_\phi \partial_z \Phi_2 \\ 
\partial_z \Phi_0 - \left( \partial_r^2 + \frac{1}{r}\partial_r + \frac{1}{r^2}\partial_\phi^2 \right) \Phi_2 \end{pmatrix} .
\label{eq:uVectorCylCoordPotentialsPhi}
\end{equation}
From Eq.~(\ref{eq:StressStrainRelNW}) and the equations in Appendix~\ref{secsub:StrainInCylindricalCoords}, one finds 
\begin{equation}
\sigma \bm{e}_r = \begin{pmatrix} \sigma_{rr} \\ \sigma_{r\phi} \\ \sigma_{rz} \end{pmatrix} = \begin{pmatrix} (2\mu + \lambda)\partial_r u_r + \lambda \left( \frac{1}{r} u_r + \frac{1}{r} \partial_\phi u_\phi + \partial_z u_z \right)  \\
\mu \left( \frac{1}{r} \partial_\phi u_r - \frac{1}{r} u_\phi + \partial_r u_\phi \right) \\ 
\mu \left( \partial_z u_r + \partial_r u_z \right) \end{pmatrix} 
\end{equation} 
for the stress related to the radial direction. For a homogeneous NW of radius $R$, the boundary conditions are
\begin{equation}
\sigma(R, \phi, z) \bm{e}_r = 0 ,
\end{equation} 
i.e., the stress tensor elements $\sigma_{rr}$, $\sigma_{r \phi}$, and $\sigma_{rz}$ must vanish at $r = R$ due to the assumption of a force-free wire surface. Using the ansatz introduced above, these boundary conditions can be written in the form
\begin{equation}
\sum_{\eta} \chi_{\eta, J} \bm{V}_\eta = \sum_{\eta} \chi_{\eta, J} \begin{pmatrix} V_{\eta,r} \\  V_{\eta,\phi} \\  V_{\eta,z} \end{pmatrix} = 0 ,
\end{equation}
where $\bm{V}_\eta$ are vectors with components $V_{\eta,r}$, $V_{\eta,\phi}$, and $V_{\eta,z}$. The boundary conditions can only be met in a nontrivial fashion (i.e., not all $\chi_{\eta, J}$ are zero) when the corresponding determinant vanishes, 
\begin{equation}
\det\bigl[\bm{V}_0, \bm{V}_1, \bm{V}_2\bigr] = \det\left[\begin{array}{ccc}
V_{0,r} & V_{1,r} & V_{2,r} \\
V_{0,\phi} & V_{1,\phi} & V_{2,\phi} \\
V_{0,z} & V_{1,z} & V_{2,z} 
\end{array}\right] = 0.
\label{eq:DeterminantHomogeneousNW}
\end{equation}
For given $n$ and $q_z$, the allowed angular frequencies $\omega$ can be found from this determinantal equation. We note, however, that a root of Eq.~(\ref{eq:DeterminantHomogeneousNW}) does not necessarily correspond to a physical solution that describes a phonon mode. The latter can be found for given $n$, $q_z$, and $\omega$ by calculating the coefficients $\chi_{\eta, J}$ from the set of boundary conditions. One of these coefficients can be chosen arbitrarily and determines the phase and amplitude of the lattice vibration. In a quantum mechanical description, this coefficient is finally obtained from the normalization condition.

There are four types of low-energetic phonon modes in a NW: one torsional ($t$; $n=0$), one longitudinal ($l$; $n=0$), and two flexural modes ($f_\pm$; $n = \pm 1$). These modes are referred to as gapless, as their angular frequencies $\omega(q_z)$ and, thus, the phonon energies $\hbar \omega(q_z)$ converge to zero when $q_z \to 0$. In the following, we summarize the dispersion relation, the displacement field, and the strain tensor elements for each of these modes. We consider the regime of lowest energy, i.e., the regime of small $q_z$ for which an expansion in $q_z r$ applies. We note that the investigated phonon modes are acoustic modes, as the atoms of a unit cell move in phase, i.e., in the same direction, in contrast to the out-of-phase movement of optical phonons where the atoms of a unit cell move in opposite directions.

\subsection{Torsional mode}
\label{secsub:TorsModeHomogeneousNW}

We start our summary with a special solution. It can easily be verified that the displacement \cite{cleland:book}
\begin{equation}
\bm{u}_{q_z t} =  c_{q_z t} r  e^{i (q_z z - \omega_{q_z t} \tau)} \bm{e}_\phi
\label{eq:uTorsHomogNWMain}
\end{equation}
meets the boundary conditions, as $0 = \sigma_{rr} = \sigma_{r\phi} = \sigma_{rz}$. The prefactor $c_{q_z t}$ is a dimensionless complex number and may be chosen arbitrarily. Furthermore, defining angular frequencies as positive, the equation of motion [Eq.~(\ref{eq:DynEqIsotrSolidMain})] is satisfied for
\begin{equation}
\omega_{q_z t}  = v_t |q_z|  = \sqrt{\frac{\mu}{\rho}} |q_z| .
\end{equation} 
Due to the displacement along $\bm{e}_\phi$, this mode is referred to as torsional ($t$), and $v_t$ is the speed of the corresponding sound wave. From Eq.~(\ref{eq:uVectorCylCoordPotentialsPhi}), it can be seen that $\bm{u}_{q_z t}$ of Eq.~(\ref{eq:uTorsHomogNWMain}) is generated via $0 = \Phi_0 = \Phi_2$ and
\begin{equation}
\Phi_1 = - \frac{c_{q_z t}}{2} r^2 e^{i (q_z z - \omega_{q_z t} \tau)} .
\end{equation} 
A special feature of this result compared to the others summarized in this work is that $\Phi_1$ does not obey the wave equation, Eq.~(\ref{eq:WaveEquationsSkalarPotsMain}). Moreover, the presented solution for the torsional mode in homogeneous NWs is exact and does not require an expansion in $q_z r$. The only nonzero strain tensor element in cylindrical coordinates is
\begin{equation}
\epsilon_{\phi z} = i \frac{c_{q_z t}}{2} q_z r e^{i (q_z z - \omega_{q_z t} \tau)} ,
\end{equation}   
and we mention that $\epsilon_{xz} = - \epsilon_{\phi z} \sin\phi$ and $\epsilon_{yz} = \epsilon_{\phi z} \cos\phi$ in cartesian coordinates.

In Sec.~\ref{secsub:TorsModeCoreShellNW}, the torsional mode in core/shell NWs is investigated with an ansatz based on Bessel functions, for which Eq.~(\ref{eq:WaveEquationsSkalarPotsMain}) is satisfied. It is therefore worth mentioning that the special solution for homogeneous NWs is obtained in the limit of a vanishing shell. We consider $0 = \Phi_0 = \Phi_2$ and
\begin{equation}
\Phi_1 =  \chi_{1,J} J_0 (\kappa_{1,J} r) e^{i(q_z z - \omega \tau)} 
\end{equation}  
as an ansatz, assuming $\kappa_{1,J} \neq 0$, i.e., $\omega \neq |q_z| \sqrt{\mu/\rho}$, which may be due to the presence of a shell. The resulting displacement function is
\begin{equation}
\bm{u} =  \chi_{1,J} \kappa_{1,J} J_1 (\kappa_{1,J} r) e^{i(q_z z - \omega \tau)} \bm{e}_\phi ,
\end{equation} 
and we note that $0 = \sigma_{rr} = \sigma_{rz}$. The arbitrary coefficient $\chi_{1,J}$ may be written as $\chi_{1,J} = 2 c_{q_z t} / \kappa_{1,J}^2$. Considering $\kappa_{1,J} r$ as a small parameter, expansion yields  
\begin{equation}
\bm{u} =  c_{q_z t} r \left[ 1 + \mathcal{O}(\kappa_{1,J}^2 r^2) \right] e^{i(q_z z - \omega \tau)} \bm{e}_\phi 
\end{equation} 
and
\begin{equation}
\sigma_{r\phi} = - \mu \frac{c_{q_z t}}{4} \kappa_{1,J}^2 r^2 \left[ 1 + \mathcal{O}(\kappa_{1,J}^2 r^2) \right] e^{i(q_z z - \omega \tau)} .
\end{equation}
For $\omega \to |q_z| \sqrt{\mu/\rho}$, i.e., $\kappa_{1,J} \to 0$, which corresponds to the limit of a vanishing shell, one finds that the boundary condition of a force-free wire surface is fulfilled due to $\sigma_{r\phi} \to 0$. As anticipated, the displacement $\bm{u}$ converges to the solution for homogeneous NWs, Eq.~(\ref{eq:uTorsHomogNWMain}).

\subsection{Longitudinal mode}
\label{secsub:LongModeHomogeneousNW}

The longitudinal and torsional modes in the NW have no angular dependence, $n = 0$. In stark contrast to the torsional mode, however, the longitudinal mode ($l$) does not lead to displacement along $\bm{e}_\phi$, and so $u_\phi = 0$. The boundary condition $\sigma_{r \phi}(R, \phi, z) = 0$ is therefore fulfilled and one may set $\chi_{1,J} = 0$ in the ansatz discussed in Sec.~\ref{secsub:PhononsEOMAnsatzBoundCond}. Analogously to Eq.~(\ref{eq:DeterminantHomogeneousNW}), the eigenfrequencies $\omega_{q_z l}$ can be calculated via the determinant of a 2$\times$2 matrix that summarizes the remaining boundary conditions. Considering angular frequencies as positive, one finds that the dominant terms of this determinant vanish for \cite{landau:elasticity, cleland:book, sirenko:pre96}
\begin{gather}
\omega_{q_z l} = v_l |q_z| = v_{l,0} |q_z| \left[ 1 + \mathcal{O}(q_z^2 R^2) \right] , \\
v_{l,0} = \sqrt{\frac{Y}{\rho}} ,
\end{gather}
where $v_l$ is the corresponding speed of sound. The properties of the longitudinal and flexural modes can conveniently be written in terms of Young's modulus and Poisson's ratio [see also Eqs.~(\ref{eq:LambdaYoungsModPoissonsRat}) and (\ref{eq:MuYoungsModPoissonsRat})]
\begin{eqnarray}
Y &=& \frac{\mu (2 \mu + 3\lambda )}{\mu + \lambda} , \\
\nu &=& \frac{\lambda}{2(\mu + \lambda)} .
\end{eqnarray}
Introducing the dimensionless $c_{q_z l}$ as an arbitrary complex prefactor, the resulting displacement vector is of the form 
\begin{equation}
\bm{u}_{q_z l} =  c_{q_z l} R \begin{pmatrix} - i \nu q_z r + \mathcal{O}(\delta^3) \\ 0 \\ 1 + \mathcal{O}(\delta^2)  \end{pmatrix} e^{i (q_z z - \omega_{q_z l} \tau)} ,
\end{equation}
with $\bm{e}_r$, $\bm{e}_\phi$, and $\bm{e}_z$ as the basis vectors. Here and in the remainder of the section, $\mathcal{O}(\delta^m)$ refers to higher-order terms of type $q_z^m R^k r^{m-k}$, where $k \geq 0$ and $m \geq k$ are integers. For the strain tensor elements in cylindrical coordinates, one obtains
\begin{gather}
\epsilon_{rr} = - i c_{q_z l} \nu q_z R \left[ 1 + \mathcal{O}(\delta^2) \right] e^{i (q_z z - \omega_{q_z l} \tau)} , \\
\epsilon_{\phi \phi} = -i c_{q_z l} \nu q_z R \left[ 1 + \mathcal{O}(\delta^2) \right] e^{i (q_z z - \omega_{q_z l} \tau)} , \\
\epsilon_{zz} = i c_{q_z l} q_z R \left[ 1 + \mathcal{O}(\delta^2) \right] e^{i (q_z z - \omega_{q_z l} \tau)} ,
\end{gather} 
and $0 = \epsilon_{r \phi} = \epsilon_{r z} = \epsilon_{\phi z}$. We note that the corrections to $\epsilon_{rr}$ and $\epsilon_{\phi\phi}$ are not identical. In cartesian coordinates, one therefore finds nonzero $\epsilon_{xy}$, with
\begin{equation}
\epsilon_{xy} =  i c_{q_z l} \sin(2\phi) \frac{\nu(1 - 2 \nu^2)}{8 (1 - \nu)} q_z^3 R r^2 \left[ 1  + \mathcal{O}(\delta^2) \right] e^{i (q_z z - \omega_{q_z l} \tau)}.
\end{equation} 
The dominant term for $\epsilon_{xx}$ and $\epsilon_{yy}$ is the same as that for $\epsilon_{rr}$ and $\epsilon_{\phi\phi}$, and for completeness we mention that $0 = \epsilon_{xz} = \epsilon_{yz}$.

\subsection{Flexural modes}
\label{secsub:FlexModesHomogeneousNW}

The flexural modes, also referred to as bending modes, comprise displacement in all three dimensions. Furthermore, the displacement is angular-dependent due to $n = \pm 1$. Considering positive $\omega$, we note that a lattice vibration of type $\exp[i(q_z z + \phi - \omega \tau)]$ cannot be written as a linear combination of those of type $\exp[i(q_z z - \phi - \omega \tau)]$, as the waves travel in opposite directions around the NW for fixed $z$. Therefore, the flexural modes $f_{+}$ and $f_{-}$, which correspond to $n = +1$ and $n= -1$, respectively, are independent. Among the gapless modes, the flexural ones are the most complicated, and neither of the $\chi_{\eta, J}$ can be set to zero in the ansatz discussed in Sec.~\ref{secsub:PhononsEOMAnsatzBoundCond}. Solving Eq.~(\ref{eq:DeterminantHomogeneousNW}) yields the parabolic dispersion relation \cite{sirenko:pre96, landau:elasticity, cleland:book} 
\begin{gather}
\omega_{q_z f} = \omega_{q_z f_+} = \omega_{q_z f_-} = \zeta_f q_z^2 = \zeta_{f,0} q_z^2 \left[ 1 + \mathcal{O}(\delta^2) \right] , \\
\zeta_{f,0} = \frac{R}{2}\sqrt{\frac{Y}{\rho}} .
\end{gather}  
The displacement vectors for $f_\pm$ can be written as  
\begin{equation}
\bm{u}_{q_z f_\pm} = c_{q_z f_\pm} R \begin{pmatrix} \mp i \pm \mathcal{O}(\delta^2) \\ 1 + \mathcal{O}(\delta^2) \\ \mp q_z r \pm \mathcal{O}(\delta^3)  \end{pmatrix} e^{i (q_z z \pm \phi - \omega_{q_z f} \tau)} ,
\end{equation}
where the components of the vector refer again to the basis $\{\bm{e}_r ,\bm{e}_\phi, \bm{e}_z \}$, and $c_{q_z f_\pm}$ are dimensionless complex prefactors. Introducing the shorthand notation
\begin{equation}
e^{(\pm)} = e^{i (q_z z \pm \phi - \omega_{q_z f} \tau)} 
\label{eq:shorthandNotPhaseFactorFlex}
\end{equation}
for convenience, the diagonal strain tensor elements for the flexural modes $f_\pm$ are
\begin{gather}
\epsilon_{rr} = \pm i c_{q_z f_\pm} \nu q_z^2 R r \left[ 1 + \mathcal{O}(\delta^2) \right] e^{(\pm)} , \\
\epsilon_{\phi \phi} = \pm i c_{q_z f_\pm} \nu q_z^2 R r \left[ 1 + \mathcal{O}(\delta^2) \right] e^{(\pm)}  , \\
\epsilon_{zz} = \mp i c_{q_z f_\pm} q_z^2 R r \left[ 1 + \mathcal{O}(\delta^2) \right] e^{(\pm)} ,
\end{gather}
and the off-diagonal ones are 
\begin{gather}
\epsilon_{r\phi} = \frac{c_{q_z f_\pm}}{48} \left[ q_z^4 R r \left(R^2 - r^2 \right) (1 - 2 \nu) + \mathcal{O}(\delta^6) \right] e^{(\pm)} , 
\label{eq:erphiFlexHomogeneous} \\
\epsilon_{rz} = \pm \frac{c_{q_z f_\pm}}{8} \left[ q_z^3 R \left( R^2 - r^2 \right) (3 + 2 \nu) + \mathcal{O}(\delta^5) \right] e^{(\pm)} , \\
\epsilon_{\phi z} = i \frac{c_{q_z f_\pm}}{8} \left[ q_z^3 R \left( R^2 (3 + 2 \nu) - r^2 (1 - 2 \nu) \right) + \mathcal{O}(\delta^5) \right] e^{(\pm)} . 
\end{gather}
From the above equations, it is evident that $\epsilon_{r\phi}$ and $\epsilon_{rz}$ vanish at $r = R$, consistent with the boundary conditions. The dominant terms of the strain components in cartesian coordinates may easily be obtained with the relations listed in Appendix~\ref{secsub:Transformation2ndRankTensors}. We note that calculation of $\epsilon_{xy}$ requires knowledge of the difference
\begin{eqnarray}
\epsilon_{rr} - \epsilon_{\phi\phi} = \pm i \frac{c_{q_z f_\pm}}{24} \left[ W_1 + \mathcal{O}(\delta^6) \right] e^{(\pm)} ,
\label{eq:errMinusephiphiFlexHomogeneous}
\end{eqnarray}
where we defined
\begin{equation}
W_1 = q_z^4 R r \left[ 2 r^2 (1 + \nu) - R^2 (1 - 2 \nu) \right] .
\end{equation}
It is worth mentioning that the listed expressions for displacement and strain do not depend on Young's modulus, whereas the dispersion relation does not depend on Poisson's ratio. The same feature is seen for the longitudinal mode.

\subsection{Normalization}
\label{secsub:NormalizeHomogeneousNW}

When quantum mechanical effects of lattice vibrations in NWs are investigated, such as, e.g., the phonon-mediated decay of spin qubits in NW QDs \cite{trif:prb08, maier:prb13}, the amplitudes of the modes are no longer arbitrary as the phonon field must be quantized \cite{stroscio:book}. Defining the time-independent displacement operator as
\begin{equation}
\bm{u}(\bm{r}) = \sum_{q_z,s} \left(a_{q_z s} \bm{u}_{q_z s}(\bm{r}, \tau=0)  + \mbox{H.c.} \right) , 
\label{eq:DisplOperatorMaintext}
\end{equation}
where $\bm{u}_{q_z s} = \bm{u}_{q_z s}(\bm{r},\tau)$ are the displacement functions discussed in this section, $s \in \{l, t, f_+, f_-\}$ indicates the mode, and ``H.c.'' stands for the Hermitian conjugate, the normalization condition for the coefficients $c_{q_z s}$ is \cite{stroscio:jap94}
\begin{equation}
\int_0^{R} dr r \hspace{0.03cm} \bm{u}_{q_z s}^{*} \cdot \bm{u}_{q_z s} = \frac{\hbar}{4 \pi L \rho \omega_{q_z s}} 
\label{eq:NormCondHomogNWsMain}
\end{equation}
in the case of homogeneous NWs. The introduced operators $a_{q_z s}^\dagger$ and $a_{q_z s}$ are the creation and annihilation operators for the phonons, and we mention that
\begin{equation}
H = \sum_{q_z,s} \hbar \omega_{q_z s} \left( a_{q_z s}^\dagger a_{q_z s} + \frac{1}{2} \right)
\end{equation} 
is the phonon Hamiltonian. As in Eq.~(\ref{eq:DisplOperatorMaintext}), the sum runs over all mode types $s$ and all wave numbers $q_z$ within the first Brillouin zone. For details, see Appendix~\ref{sec:NormalizationNWs} (wherein $\bm{u}_{q_z s}$ is written as $c_{q_z s} \bm{u}_{q_z s}$ for illustration purposes).

From Eq.~(\ref{eq:NormCondHomogNWsMain}) and the derived expressions for $\bm{u}_{q_z s}$, we calculate the normalization condition for the coefficients $c_{q_z s}$ in leading order of $q_z R$. For the torsional, longitudinal, and flexural modes, respectively, one finds
\begin{gather}
|c_{q_z t}|^2 = \frac{\hbar}{\pi L R^4 \rho v_t |q_z| } , \\
|c_{q_z l}|^2 = \frac{\hbar}{2 \pi L R^4 \rho v_{l,0} |q_z| } \left[ 1 + \mathcal{O}(\delta^2) \right] , \\
|c_{q_z f_{\pm}}|^2 = \frac{\hbar}{4 \pi  L R^4  \rho \zeta_{f,0} q_z^2 } \left[1 + \mathcal{O}(\delta^2) \right] .
\end{gather}

\section{Phonons in core/shell nanowires}
\label{sec:PhononsCoreShellWireMain}

We now extend the analysis of the previous section to the more complicated case of core/shell NWs. Considering the complexity of the system, the resulting formulas are surprisingly simple, and so we believe that our results will prove very helpful in future investigations that involve acoustic phonons in core/shell NWs.

\subsection{Ansatz and boundary conditions}
\label{secsub:PhononsAnsatzBCondCoreShell}

We assume pseudomorphic growth and start from a core/shell NW that is statically strained. The static strain in the NW was calculated in Sec.~\ref{sec:StaticStrainMain} and ensures that the surface is free of forces, that the core-shell interface is coherent, and that the stress at the interface is continuous. The dynamical displacement field of the lattice vibrations discussed in this section describes the displacement from this statically strained configuration. As the lattice mismatch $|\epsilon_0| \ll 1$ is small, the static and dynamical displacement fields can be considered as independent and add linearly in good approximation (analogous for the stress and strain) \cite{cleland:book}. 

\begin{figure}[t]
\begin{center}
\includegraphics[width=0.85\linewidth]{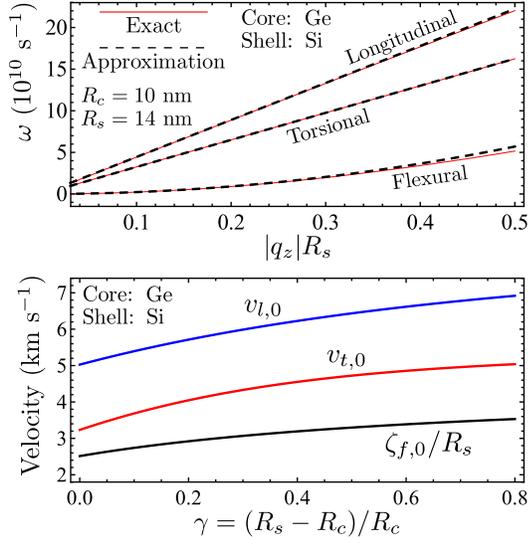}
\caption{Dispersion relation of gapless phonon modes in Ge/Si core/shell NWs. Top: The thin red lines are numerically calculated roots of the determinant that comprises the nine boundary conditions described in Sec.~\ref{secsub:PhononsAnsatzBCondCoreShell}. For the parameters in Appendix~\ref{sec:ParametersGeSiCoreShell} and an assumed core (shell) radius of $R_c = \mbox{10 nm}$ ($R_s = \mbox{14 nm}$), phonons with gapped spectra were found at $\omega > 5\times 10^{11}\mbox{s}^{-1}$, i.e., $\hbar \omega > 0.3\mbox{ meV}$. The dashed black lines correspond to $\omega = v_{t,0} |q_z|$, $\omega = v_{l,0} |q_z|$, and $\omega = \zeta_{f,0} q_z^2$, respectively, and agree well with the exact result even at relatively large $q_z$. Bottom: The mode velocities $v_{t,0}$ [torsional, Eq.~(\ref{eq:vt0coreshell})], $v_{l,0}$ [longitudinal, Eq.~(\ref{eq:vl0coreshell})], and $\zeta_{f,0}/R_s$ [flexural, Eq.~(\ref{eq:zetaf0coreshell})] are plotted as a function of the relative shell thickness~$\gamma$. }
\label{fig:DispersionRelations}
\end{center}
\end{figure}

The dynamical displacement $\bm{u}^{c,s}$ and the stress tensor $\sigma^{c,s}$ for core ($c$; $0 \leq r \leq R_c$) and shell ($s$; $R_c \leq r \leq R_s$), respectively, are calculated with the ansatz introduced in Sec.~\ref{secsub:PhononsEOMAnsatzBoundCond} \cite{nishiguchi:prb94}. Suitable functions $\Phi_\eta^{c,s}$ are
\begin{gather}
\Phi_\eta^{c} = \chi_{\eta, J}^{c} J_n (\kappa_{\eta,J}^{c} r) e^{i(q_z z + n \phi - \omega \tau)} , \\
\Phi_\eta^{s} = \left[ \chi_{\eta, J}^{s} J_n (\kappa_{\eta,J}^{s} r) + \chi_{\eta, Y}^{s} Y_n (\kappa_{\eta,Y}^{s} r) \right] e^{i(q_z z + n \phi - \omega \tau)} ,
\end{gather} 
where
\begin{gather}
( \kappa_{\eta,J}^{c} )^2 = \frac{\rho_c \omega^2}{\mu_c + \delta_{\eta,0}(\mu_c + \lambda_c)}  - q_z^2 , \\
(\kappa_{\eta,J}^{s})^2 = (\kappa_{\eta,Y}^{s})^2 = \frac{\rho_s \omega^2}{\mu_s + \delta_{\eta,0}(\mu_s + \lambda_s)}  - q_z^2 .
\end{gather}
The boundary conditions are the same as for the static strain and can be summarized as 
\begin{eqnarray}
\bm{u}^c(R_c, \phi, z) &=& \bm{u}^s(R_c, \phi, z) , \\
\sigma^{c}(R_c, \phi, z) \bm{e}_r &=& \sigma^{s}(R_c, \phi, z) \bm{e}_r , \\
\sigma^{s}(R_s, \phi, z) \bm{e}_r &=& 0.
\end{eqnarray}
For given mode type and wave number $q_z$, these boundary conditions determine the eigenfrequency and the set of coefficients $\{ \chi_{\eta, J}^{c,s}, \chi_{\eta, Y}^{s} \}$. As in the case of homogeneous NWs, one of the coefficients may be chosen arbitrarily and quantifies the amplitude and phase of the lattice vibration. Analogous to Eq.~(\ref{eq:DeterminantHomogeneousNW}), the eigenfrequency can be calculated by solving a determinantal equation that comprises the boundary conditions (see also Fig.~\ref{fig:DispersionRelations}).

In order to derive algebraic expressions, we consider again the regime of lowest energy (small $q_z$) for which an expansion in $q_z r$ applies. In this section, higher-order contributions denoted by $\mathcal{O}(\delta^m)$ refer to corrections of type $q_z^m R_s^j R_c^k r^l$, where $j$, $k$, $l$, and $m = j + k + l$ are integers. We note that $l<0$ is allowed in the shell due to the Bessel functions of the second kind. While we list the dominant terms of the displacement field for both the core and the shell, the phonon-based strain tensor is provided in detail for the core only. This is typically sufficient, as qubit states, for instance, are usually confined therein.

\subsection{Torsional mode}
\label{secsub:TorsModeCoreShellNW}

As expected from symmetry considerations, it turns out that the main features of the gapless phonon modes remain unchanged when the NW is surrounded by a shell. For instance, the torsional mode has no angular dependence ($n=0$) and involves displacement along $\bm{e}_\phi$ only. Thus, one obtains $0 = \chi_{0, J}^{c,s} = \chi_{2, J}^{c,s} = \chi_{0, Y}^{s} = \chi_{2, Y}^{s}$, and the eigenfrequency $\omega_{q_z t}$ can be calculated via the determinant of a 3$\times$3 matrix that contains the three remaining boundary conditions. We find
\begin{gather}
\omega_{q_z t} = v_t |q_z| = v_{t,0} |q_z| \left[ 1 + \mathcal{O}(\delta^2) \right] , \\
v_{t,0} = \sqrt{\frac{\mu_c R_c^4 + \mu_s (R_s^4 - R_c^4) }{\rho_c R_c^4 + \rho_s (R_s^4 - R_c^4)}} ,
\end{gather}
and note that $v_{t,0}$, in fact, is a function of the relative shell thickness. Defining
\begin{equation}
\mathring{\gamma} = \widetilde{\gamma}^2 + 2 \widetilde{\gamma} = \frac{R_s^4 - R_c^4}{R_c^4} 
\label{eq:gammaDefinitionRadiusToFour}
\end{equation}
analogously to Eqs.~(\ref{eq:gammaRelShThDefinition}) and (\ref{eq:gammaTildeDefinition}), the mode velocity $v_{t,0}$ has the remarkably simple form  
\begin{equation}
v_{t,0} = \sqrt{\frac{\mu_c + \mu_s \mathring{\gamma}}{\rho_c + \rho_s \mathring{\gamma}}} .
\label{eq:vt0coreshell}
\end{equation}
The displacement field is
\begin{eqnarray}
\bm{u}_{q_z t}^{c} &=&  c_{q_z t} r \left[ 1 + \mathcal{O}(\delta^2) \right]    e^{i (q_z z - \omega_{q_z t} \tau)} \bm{e}_\phi , \\
\bm{u}_{q_z t}^{s} &=&  c_{q_z t} r \left[ 1 + \mathcal{O}(\delta^2) \right]    e^{i (q_z z - \omega_{q_z t} \tau)} \bm{e}_\phi ,
\end{eqnarray} 
i.e., the dominant term for core and shell is the same. The strain tensor elements for the core are
\begin{gather}
\epsilon_{r \phi}^{c} = \frac{c_{q_z t}}{8} \left[ q_z^2 r^2 \left(1 - v_{t,0}^2 \frac{\rho_c}{\mu_c} \right) + \mathcal{O}(\delta^4) \right] e^{i (q_z z - \omega_{q_z t} \tau)} , \\
\epsilon_{\phi z}^{c} = i \frac{c_{q_z t}}{2}  \left[ q_z r + \mathcal{O}(\delta^3) \right] e^{i (q_z z - \omega_{q_z t} \tau)} ,
\end{gather}   
and $0 = \epsilon_{rr}^{c} = \epsilon_{\phi\phi}^{c} = \epsilon_{zz}^{c} = \epsilon_{rz}^{c}$. We emphasize that $\epsilon_{r \phi}^{c}$ is nonzero, in stark contrast to $\epsilon_{r \phi}$ in homogeneous NWs. The strain components in cartesian coordinates, among which only $\epsilon_{zz}^{c}$ is zero, can be calculated with the above-mentioned $\epsilon_{r \phi}^{c}$ and $\epsilon_{\phi z}^{c}$ and the relations in Appendix~\ref{secsub:Transformation2ndRankTensors}.

\begin{figure}[t]
\begin{center}
\includegraphics[width=0.85\linewidth]{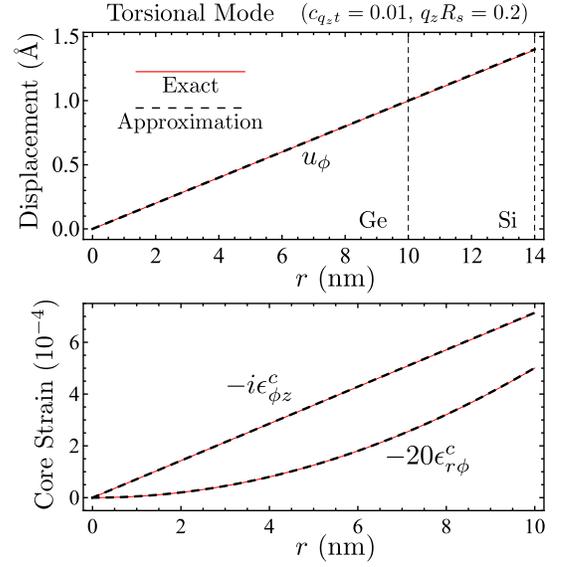}
\caption{Radial dependence of the displacement field (top) and core strain (bottom) in cylindrical coordinates due to the torsional phonon mode in a Ge/Si NW with core radius $R_c = \mbox{10 nm}$ and shell radius $R_s = \mbox{14 nm}$. Thin red lines correspond to the exact, numerical solution of the ansatz described in Sec.~\ref{secsub:PhononsAnsatzBCondCoreShell}, while dashed black lines are calculated with the algebraic expressions listed in Sec.~\ref{secsub:TorsModeCoreShellNW}. Besides the parameters for Ge and Si in the text (see also Appendix~\ref{sec:ParametersGeSiCoreShell}), we use $c_{q_z t} = 0.01$ and $q_z = 0.2/R_s$. The global phase factor $\exp[i(q_z z - \omega_{q_z t} \tau)]$ is set to 1. Excellent agreement between the exact and approximate solutions is found even when $q_z$ is relatively large. We note that $0 = u_r = u_z$ and $0 = \epsilon_{rr}^{c} = \epsilon_{\phi\phi}^{c} = \epsilon_{zz}^{c} = \epsilon_{rz}^{c}$.}
\label{fig:TorsionalModeResults}
\end{center}
\end{figure} 

In Fig.~\ref{fig:DispersionRelations} (top), we illustrate that the derived formula for the velocity $v_{t,0}$ is consistent with exact, numerical solutions of the underlying model (Sec.~\ref{secsub:PhononsAnsatzBCondCoreShell}). This also holds for the dispersion relation of the longitudinal and flexural modes that we investigate next. The dependence of the different phonon velocities on the relative shell thickness of a Ge/Si core/shell NW is shown in Fig.~\ref{fig:DispersionRelations} (bottom). A comparison between exact results and the above-listed expressions for the displacement and strain caused by torsional lattice vibrations is provided in Fig.~\ref{fig:TorsionalModeResults}.

\subsection{Longitudinal mode}
\label{secsub:LongModeCoreShellNW}

\begin{figure}[t]
\begin{center}
\includegraphics[width=0.85\linewidth]{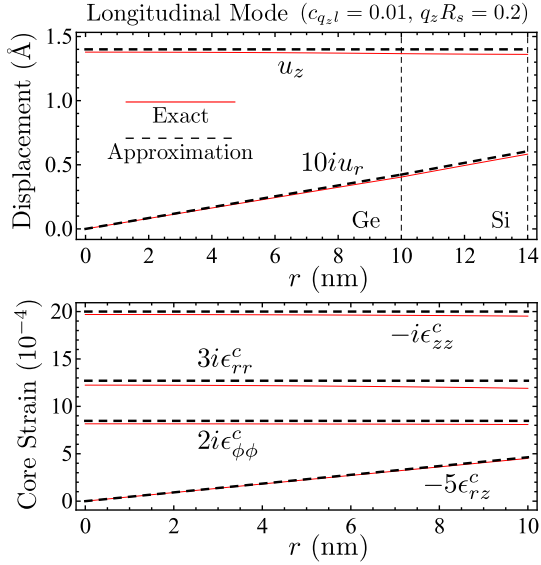}
\caption{Comparison between exact results and derived formulas for the longitudinal phonon mode, analogous to Fig.~\ref{fig:TorsionalModeResults}. Good agreement is found for all components at $q_z = 0.2/R_s$ assumed here, and the quality of the approximation increases with decreasing $|q_z|$. The longitudinal mode features $u_\phi = 0$ and $0 = \epsilon_{r\phi}^{c} = \epsilon_{\phi z}^{c}$. For details, see Sec.~\ref{secsub:LongModeCoreShellNW}.}
\label{fig:LongitudinalModeResults}
\end{center}
\end{figure}

In the ansatz for the longitudinal mode, we set $n = 0$ and $0 = \chi_{1,J}^{c,s} = \chi_{1,Y}^{s}$. From the boundary conditions, we obtain
\begin{gather}
\omega_{q_z l} = v_l |q_z| = v_{l,0} |q_z| \left[ 1 + \mathcal{O}(\delta^2) \right] , \\
v_{l,0} = \sqrt{\frac{Y_c G_0 + Y_s G_1 \widetilde{\gamma}^2 + Y_c G_2 \widetilde{\gamma}}{(G_0 + G_1 \widetilde{\gamma})(\rho_c + \rho_s \widetilde{\gamma})}} ,
\label{eq:vl0coreshell}
\end{gather}
where we introduced
\begin{gather}
G_0 = 2 Y_c (1 - \nu_s^2) , \\
G_1 = Y_c (1 + \nu_s) + Y_s (1 - \nu_c - 2 \nu_c^2) , \\
G_2 = Y_c (1 + \nu_s) + Y_s (3 - \nu_c - 4 \nu_c \nu_s) 
\end{gather}
for convenience. The resulting displacement field $\bm{u}_{q_z l}$ is
\begin{equation}
\bm{u}_{q_z l}^c =  c_{q_z l} R_s \begin{pmatrix} - i \frac{\nu_c G_0 + G_3 \widetilde{\gamma}}{G_0 + G_1 \widetilde{\gamma}} q_z r + \mathcal{O}(\delta^3) \\ 0 \\ 1 + \mathcal{O}(\delta^2)  \end{pmatrix} e^{i (q_z z - \omega_{q_z l} \tau)} 
\end{equation}
within the core and
\begin{equation}
\bm{u}_{q_z l}^s =  c_{q_z l} R_s \begin{pmatrix} - i \left[ \frac{(G_4 + \nu_s G_1 \widetilde{\gamma})q_z r}{G_0 + G_1 \widetilde{\gamma}} + \frac{G_5 q_z R_s^2 }{(G_0 + G_1 \widetilde{\gamma}) r } \right] + \mathcal{O}(\delta^3) \\ 0 \\ 1 + \mathcal{O}(\delta^2)  \end{pmatrix} e^{i (q_z z - \omega_{q_z l} \tau)} 
\end{equation}
within the shell, where
\begin{gather}
G_3 = Y_c (1 + \nu_s) \nu_c + Y_s (1 - \nu_c - 2 \nu_c^2) \nu_s , \\
G_4 = Y_c (1 + \nu_s)(\nu_c + \nu_s - 2 \nu_c \nu_s) , \\
G_5 = Y_c (1 + \nu_s)(\nu_c - \nu_s) .
\end{gather}
An example based on Ge/Si NWs for the displacement and the strain discussed below is shown in Fig.~\ref{fig:LongitudinalModeResults}, where we also provide a comparison with the exact solution. 

The diagonal strain tensor elements for the core are similar to those for homogeneous NWs,
\begin{gather}
\epsilon_{rr}^{c} = - i c_{q_z l}  \frac{\nu_c G_0 + G_3 \widetilde{\gamma}}{G_0 + G_1 \widetilde{\gamma}} q_z R_s \left[ 1 + \mathcal{O}(\delta^2) \right] e^{i (q_z z - \omega_{q_z l} \tau)} , \\
\epsilon_{\phi\phi}^{c} = - i c_{q_z l}  \frac{\nu_c G_0 + G_3 \widetilde{\gamma}}{G_0 + G_1 \widetilde{\gamma}} q_z R_s \left[ 1 + \mathcal{O}(\delta^2) \right] e^{i (q_z z - \omega_{q_z l} \tau)} , \\
\epsilon_{zz}^{c} = i c_{q_z l} q_z R_s \left[ 1 + \mathcal{O}(\delta^2) \right] e^{i (q_z z - \omega_{q_z l} \tau)} . 
\end{gather}
Furthermore, we note that $0 = \epsilon_{r\phi}^{c} = \epsilon_{\phi z}^{c}$ as the longitudinal mode is independent of $\phi$ and does not provide displacement along $\bm{e}_\phi$. In stark contrast to homogeneous NWs, however, we find nonzero $\epsilon_{rz}^{c}$,
\begin{eqnarray}
\epsilon_{rz}^{c} &=& c_{q_z l} (1 + \nu_c) q_z^2 R_s r \left[ 1 + \mathcal{O}(\delta^2) \right] e^{i (q_z z - \omega_{q_z l} \tau)} \nonumber \\
& &\times \frac{Y_c (G_0 \rho_s - G_6 \rho_c) \widetilde{\gamma} + (Y_c G_7 \rho_s - Y_s G_1 \rho_c)\widetilde{\gamma}^2 }{2 Y_c (G_0 + G_1 \widetilde{\gamma})(\rho_c + \rho_s \widetilde{\gamma})} , \hspace{0.3cm}
\end{eqnarray}
with 
\begin{gather}
G_6 = 2 Y_s (1 - \nu_c \nu_s ) , \\
G_7 = Y_c (1 + \nu_s) + Y_s (1 - \nu_c - 2 \nu_c \nu_s) .
\end{gather}
The dominant term for $\epsilon_{rr}^c$ and $\epsilon_{\phi\phi}^c$ is identical to that for $\epsilon_{xx}^c$ and $\epsilon_{yy}^c$, and the off-diagonal strain components in cartesian coordinates may be calculated via $\epsilon_{xz}^c = \epsilon_{rz}^c \cos\phi$, $\epsilon_{yz}^c = \epsilon_{rz}^c \sin\phi$, and $\epsilon_{xy}^c = (\epsilon_{rr}^c - \epsilon_{\phi\phi}^c) \sin\phi \cos\phi$.
The resulting formula for $\epsilon_{xy}^c$ involves higher-order corrections to $\epsilon_{rr}^c$ and $\epsilon_{\phi\phi}^c$ and is too lengthy to be displayed here in its entirety. Nevertheless, we provide an approximation that applies to the case of a very thin shell ($\widetilde{\gamma} \ll 1$), 
\begin{eqnarray}
\epsilon_{xy}^{c} &\simeq& i c_{q_z l} \sin(2\phi) q_z^3 R_s r^2 \left[ 1 + \mathcal{O}(\delta^2) \right] e^{i (q_z z - \omega_{q_z l} \tau)}  
\nonumber \\
& &\times \left[\frac{\nu_c (1 - 2 \nu_c^2)}{8 (1 - \nu_c)} + \frac{\widetilde{\gamma} (1 + \nu_c)}{16 (1 - \nu_c)}\left(\frac{G_8}{G_0} - \frac{\rho^\prime}{\rho_c} \right) \right] . 
\label{eq:exyLongModeCoreShellSmallGamma}
\end{eqnarray}
In the opposite limit of a very thick shell, we find
\begin{eqnarray}
\lim_{\gamma \to \infty} \epsilon_{xy}^{c} &=& i c_{q_z l} \sin(2\phi) q_z^3 R_s r^2 \left[ 1 + \mathcal{O}(\delta^2) \right] e^{i (q_z z - \omega_{q_z l} \tau)}    \nonumber \\
& &\times \frac{1}{16 G_1}\left(\frac{Y_s (1 + \nu_c) \rho_c}{Y_c (1 - \nu_c) \rho_s} G_9 - G_{10} \right) . 
\label{eq:exyLongModeCoreShellInfiniteGamma}
\end{eqnarray}
In Eq.~(\ref{eq:exyLongModeCoreShellSmallGamma}),
\begin{eqnarray}
G_8 = 2 Y_s &\Bigl[& 1 + \nu_s + \nu_c (1-3\nu_s) \nonumber \\  
& & + 2 \nu_c^2 (2 \nu_c - 1)(3\nu_s + 1) - 8 \nu_c^4 \Bigr] 
\end{eqnarray}
and 
\begin{gather}
\rho^\prime = \rho_s ( 1 + 2 \nu_c - 4 \nu_c^2 ) ,
\end{gather}
while
\begin{eqnarray}
G_9 &=& Y_c ( 1 + 2 \nu_c - 4 \nu_c^2 ) ( 1 + \nu_s ) \nonumber \\
& & + Y_s ( 1 - \nu_c - 2 \nu_c^2 ) ( 1 + 2 \nu_s - 4 \nu_c \nu_s ) , \\
G_{10} &=& Y_c ( 1 + 2 \nu_c ) ( 1 + \nu_s ) + Y_s ( 1 + \nu_c ) ( 1 - 4 \nu_c \nu_s ) \hspace{0.3cm}
\end{eqnarray} 
in Eq.~(\ref{eq:exyLongModeCoreShellInfiniteGamma}).

\subsection{Flexural modes}
\label{secsub:FlexModeCoreShellNW}

The calculation for the flexural modes in core/shell NWs is most complicated as neither of the coefficients $\chi_{\eta,J}^{c,s}$ and $\chi_{\eta,Y}^{s}$ is zero. Furthermore, the flexural modes have an angular dependence due to $n = \pm 1$. Despite this complexity, the resulting formulas are relatively simple and can be written in a compact form. By solving the determinantal equation that comprises the nine boundary conditions, we find the dispersion relation
\begin{gather}
\omega_{q_z f} = \omega_{q_z f_+} = \omega_{q_z f_-} = \zeta_f q_z^2 = \zeta_{f,0} q_z^2 \left[ 1 + \mathcal{O}(\delta^2) \right] , \\
\zeta_{f,0} = \frac{R_s}{2}\sqrt{\frac{Y_c K_0 + Y_s K_1 \mathring{\gamma}^2 + Y_c K_2 \mathring{\gamma}}{(1 + \widetilde{\gamma})(K_0 + K_1 \mathring{\gamma})(\rho_c + \rho_s \widetilde{\gamma})}}  ,
\label{eq:zetaf0coreshell}
\end{gather} 
where
\begin{gather}
K_0 = 4 Y_c (1 - \nu_s^2) , \\
K_1 = Y_c (1 + \nu_s) + Y_s (3 - \nu_c - 4 \nu_c^2) , \\
K_2 = Y_c (1 + \nu_s) + Y_s (7 - \nu_c - 8 \nu_c \nu_s) , 
\end{gather}
and $\mathring{\gamma} = \widetilde{\gamma}^2 + 2\widetilde{\gamma}$ has been introduced in Eq.~(\ref{eq:gammaDefinitionRadiusToFour}). We note that $R_s/\sqrt{1 + \widetilde{\gamma}} = R_s/(1 + \gamma) = R_c$ in the expression for $\zeta_{f,0}$ may be substituted by the core radius. Remarkably, the dominant terms of the displacement field in both core and shell turn out to be equivalent to those for a homogeneous NW. Referring again to the basis vectors $\bm{e}_r$, $\bm{e}_\phi$, and $\bm{e}_z$, we obtain 
\begin{equation}
\bm{u}_{q_z f_\pm}^{c} = c_{q_z f_\pm} R_s \begin{pmatrix} \mp i \pm \mathcal{O}(\delta^2) \\ 1 + \mathcal{O}(\delta^2) \\ \mp q_z r \pm \mathcal{O}(\delta^3)  \end{pmatrix} e^{(\pm)} 
\end{equation}
for the core, and the formally identical result
\begin{equation}
\bm{u}_{q_z f_\pm}^{s} = c_{q_z f_\pm} R_s \begin{pmatrix} \mp i \pm \mathcal{O}(\delta^2) \\ 1 + \mathcal{O}(\delta^2) \\ \mp q_z r \pm \mathcal{O}(\delta^3)  \end{pmatrix} e^{(\pm)} 
\end{equation}
for the shell. The shorthand notation $e^{(\pm)}$ for the phase factor has been introduced in Eq.~(\ref{eq:shorthandNotPhaseFactorFlex}).

\begin{figure}[t]
\begin{center}
\includegraphics[width=0.85\linewidth]{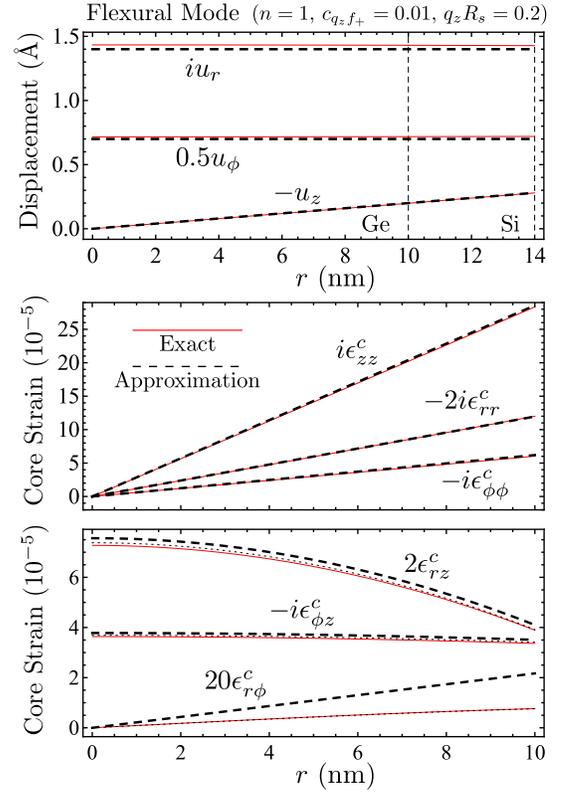}
\caption{Comparison between exact and approximate solutions for a flexural phonon mode ($n=1$), analogous to Figs.~\ref{fig:TorsionalModeResults} and \ref{fig:LongitudinalModeResults}. Assuming again $q_z = 0.2/R_s$, the top and middle figures exhibit very good agreement for the displacement and the diagonal strain components, respectively. In the bottom figure, $\epsilon_{rz}^{c}$ and $\epsilon_{\phi z}^{c}$ are well approximated by Eqs.~(\ref{eq:erzcFlexInfiniteShell}) and (\ref{eq:ephizcFlexInfiniteShell}) (dashed black lines, $\widetilde{\gamma} \to \infty$). Equations (\ref{eq:erzcFlexVeryThinShell}) and (\ref{eq:ephizcFlexVeryThinShell}) (not plotted, $\widetilde{\gamma} \ll 1$) yield a slightly worse but still good approximation. We note that the quantitative agreement can be improved by taking $\widetilde{\gamma} \sim 1$ of the studied NW fully into account in the terms of order $c_{q_z f_\pm}\mathcal{O}(\delta^3)$ (thin dotted lines for $\epsilon_{rz}^{c}$ and $\epsilon_{\phi z}^{c}$, expressions too lengthy for text). For $\epsilon_{r\phi}^{c}$, the observed deviation between Eq.~(\ref{eq:erphicFlexCoreShell}) (dashed black line) and the exact result decreases rapidly with decreasing $|q_z|$, as expected. We illustrate that corrections of order $c_{q_z f_\pm}\mathcal{O}(\delta^4)$ become important in the considered example by taking them into account (thin dotted line for $\epsilon_{r\phi}^{c}$, expressions too long for text). For details, see Sec.\ \ref{secsub:FlexModeCoreShellNW}. }
\label{fig:FlexuralModeResults}
\end{center}
\end{figure}

With the definitions
\begin{gather}
K_3 = Y_c (1 + \nu_s) \nu_c + Y_s (1 + \nu_c) (\nu_s + 2 \nu_c - 4 \nu_c \nu_s) , \\
K_4 = Y_c (1 + \nu_s) \nu_c + Y_s ( 3 - \nu_c - 4 \nu_c^2) \nu_s , 
\end{gather}
the diagonal core strain components due to the flexural modes $f_{\pm}$ are
\begin{gather}
\epsilon_{rr}^c = \pm i c_{q_z f_\pm} \frac{\nu_c K_0 + K_3 \mathring{\gamma}}{K_0 + K_1 \mathring{\gamma}}  q_z^2 R_s r \left[ 1 + \mathcal{O}(\delta^2) \right] e^{(\pm)} , \\
\epsilon_{\phi\phi}^c = \pm i c_{q_z f_\pm} \frac{\nu_c K_0 + K_4 \mathring{\gamma}}{K_0 + K_1 \mathring{\gamma}} q_z^2 R_s r \left[ 1 + \mathcal{O}(\delta^2) \right] e^{(\pm)}  , \\
\epsilon_{zz}^c = \mp i c_{q_z f_\pm} q_z^2 R_s r \left[ 1 + \mathcal{O}(\delta^2) \right] e^{(\pm)} .
\end{gather}
Thus, in contrast to the case of homogeneous NWs, the leading-order terms for $\epsilon_{rr}^c$ and $\epsilon_{\phi\phi}^c$ differ when Poisson's ratios $\nu_c$ and $\nu_s$ are different,
\begin{equation}
\epsilon_{rr}^c - \epsilon_{\phi\phi}^c = \pm i c_{q_z f_\pm} \left[ \frac{2 Y_s (1 + \nu_c) (\nu_c - \nu_s) \mathring{\gamma}}{K_0 + K_1 \mathring{\gamma}}  q_z^2 R_s r  + \mathcal{O}(\delta^4) \right] e^{(\pm)} .
\label{eq:errcMinusephiphicFlexCoreShell}
\end{equation}
Similarly, the off-diagonal strain tensor element
\begin{equation}
\epsilon_{r\phi}^c = c_{q_z f_\pm} \left[ \frac{Y_s (1 + \nu_c) (\nu_s - \nu_c) \mathring{\gamma}}{K_0 + K_1 \mathring{\gamma}}  q_z^2 R_s r  + \mathcal{O}(\delta^4) \right] e^{(\pm)} 
\label{eq:erphicFlexCoreShell}
\end{equation}
exhibits a new term that vanishes in homogeneous NWs or when $\nu_c = \nu_s$. The expressions for $\epsilon_{rz}^c$ and $\epsilon_{\phi z}^c$ are rather lengthy and therefore cannot be provided here completely. When $\widetilde{\gamma} \ll 1$, they are well approximated by
\begin{eqnarray}
\epsilon_{rz}^{c} &\simeq& \pm c_{q_z f_\pm} q_z R_s  e^{(\pm)}  
\label{eq:erzcFlexVeryThinShell} \\
& &\times \left[q_z^2 \frac{(R_s^2 - r^2) \nu^\prime }{8} + \widetilde{\gamma} q_z^2 \frac{R_s^2 K_5 + r^2 K_6}{2 K_0} + \mathcal{O}(\delta^4)  \right] , \nonumber  \\
\epsilon_{\phi z}^{c} &\simeq& i c_{q_z f_\pm} q_z R_s  e^{(\pm)} 
\label{eq:ephizcFlexVeryThinShell} \\
& &\times \left[q_z^2 \frac{R_s^2 \nu^\prime - r^2 (1 - 2 \nu_c)}{8} + \widetilde{\gamma} q_z^2 \frac{R_s^2 K_5 + r^2 K_7}{2 K_0} + \mathcal{O}(\delta^4) \right] ,  \nonumber 
\end{eqnarray} 
where 
\begin{eqnarray}
K_5 &=& Y_s (1 + \nu_c) (3 - \nu_s + 4 \nu_c^2 - 6 \nu_c \nu_s) \nonumber \\
& & - Y_c (3 + 2 \nu_c)  (1 - \nu_s^2 )  
\end{eqnarray}
and 
\begin{gather}
K_6 = 2 Y_s (1 + 3 \nu_c + 2 \nu_c^2) (\nu_s - \nu_c) , \\
K_7 = 2 Y_s (3 + \nu_c - 2 \nu_c^2) (\nu_s - \nu_c) , \\
\nu^\prime = 3 + 2 \nu_c .
\end{gather}
In the opposite regime, i.e., $\widetilde{\gamma} \gg 1$, the results for $\epsilon_{rz}^c$ and $\epsilon_{\phi z}^c$ converge to 
\begin{gather}
\lim_{\gamma \to \infty} \epsilon_{rz}^{c} = \pm c_{q_z f_\pm} \left[ q_z^3 R_s \left(\frac{R_s^2 K_8 }{8 K_9} - \frac{r^2 K_{10}}{8 K_1} \right) + \mathcal{O}(\delta^5) \right] e^{(\pm)} , \label{eq:erzcFlexInfiniteShell} \\
\lim_{\gamma \to \infty} \epsilon_{\phi z}^{c} = i c_{q_z f_\pm} \left[q_z^3 R_s \left(\frac{R_s^2 K_8 }{8 K_9} - \frac{r^2 K_{11}}{8 K_1} \right)  + \mathcal{O}(\delta^5) \right] e^{(\pm)} , \label{eq:ephizcFlexInfiniteShell}
\end{gather} 
where we defined
\begin{gather}
K_8 = 2 Y_s (1 + \nu_c) (3 + 2 \nu_s) , \\
K_9 = Y_c (1 + \nu_s) + Y_s (1 + \nu_c) ,
\end{gather}
and
\begin{eqnarray}
K_{10} &=& Y_c (3 + 2 \nu_c) (1 + \nu_s) \nonumber \\
& & + Y_s (1 + \nu_c) (9 - 2 \nu_c - 4 \nu_s - 8 \nu_c \nu_s) , \\
K_{11} &=& Y_c (1 - 2 \nu_c) (1 + \nu_s) \nonumber \\
& & + Y_s (1 + \nu_c) (3 + 2 \nu_c - 12 \nu_s + 8 \nu_c \nu_s) .
\end{eqnarray}
The strain tensor elements in cartesian coordinates may again be calculated with the relations listed in Appendix~\ref{secsub:Transformation2ndRankTensors}. The radial dependence of the displacement field and the core strain of a flexural lattice vibration is plotted in Fig.~\ref{fig:FlexuralModeResults}, confirming that our formulas are consistent with exact solutions.

As mentioned before, the new terms in Eqs.~(\ref{eq:errcMinusephiphicFlexCoreShell}) and (\ref{eq:erphicFlexCoreShell}) vanish when $R_s \to R_c$ or $\nu_s \to \nu_c$. Therefore, the higher-order contributions of order $c_{q_z f_\pm} \mathcal{O}(\delta^4)$ may become important for the calculation of $\epsilon_{r\phi}^{c}$ and $\epsilon_{xy}^{c}$, particularly when $q_z R_s$ is rather large and $\gamma$ and $\nu_s - \nu_c$ are small [see also Fig.~\ref{fig:FlexuralModeResults} (bottom)]. The expressions for these higher-order corrections in core/shell NWs are too lengthy to be displayed here. However, if needed, they can be approximated via the formulas that we provide in Eqs.~(\ref{eq:erphiFlexHomogeneous}) and (\ref{eq:errMinusephiphiFlexHomogeneous}) for homogeneous NWs.

\subsection{Normalization}
\label{secsub:NormalizeCoreShellNW}

The normalization condition of Eq.~(\ref{eq:NormCondHomogNWsMain}) applies only to the special case of homogeneous NWs with cylindrical symmetry. In the more general case of cylindrically symmetric core/shell and core/multishell NWs, the normalization condition reads
\begin{equation}
\int_0^{R_{\rm tot}} dr r \rho(r) \hspace{0.03cm} \bm{u}_{q_z s}^{*} \cdot \bm{u}_{q_z s} = \frac{\hbar}{4 \pi L \omega_{q_z s}} ,
\label{eq:NormCondCoreShellNWsMain}
\end{equation}
where $\rho(r)$ is the radially dependent density, $s \in \{l, t, f_+, f_-\}$ is the mode type, and $R_{\rm tot}$ is the total radius of the NW, i.e., the radius of the outermost shell. Considering core/shell NWs, we therefore obtain the normalization conditions
\begin{gather}
|c_{q_z t}|^2 = \frac{\hbar (1 + \mathring{\gamma}) }{\pi L R_s^4 (\rho_c + \rho_s \mathring{\gamma}) v_{t,0} |q_z| } \left[ 1 + \mathcal{O}(\delta^2) \right] , \\
|c_{q_z l}|^2 = \frac{\hbar (1 + \widetilde{\gamma}) }{2 \pi L R_s^4 (\rho_c + \rho_s \widetilde{\gamma}) v_{l,0} |q_z| } \left[ 1 + \mathcal{O}(\delta^2) \right] , \\
|c_{q_z f_{\pm}}|^2 = \frac{\hbar (1 + \widetilde{\gamma}) }{4 \pi  L R_s^4  (\rho_c + \rho_s \widetilde{\gamma}) \zeta_{f,0} q_z^2 } \left[1 + \mathcal{O}(\delta^2) \right] 
\end{gather}
for the coefficients of the phonons investigated in this section. Details about the derivation of Eq.~(\ref{eq:NormCondCoreShellNWsMain}) are provided in Appendixes~\ref{sec:PhononsBulk} and \ref{sec:NormalizationNWs}.

\subsection{Limits of vanishing and infinite shell}
\label{secsub:Limits}

Our results for phonons in core/shell NWs are fully consistent with those in Sec.~\ref{sec:PhononsBareWireMain} for homogeneous NWs. For instance, when $R_s \to R_c$, i.e., $\gamma \to 0$, the expressions for the mode velocities and for the displacement and strain in the core converge exactly to those for a homogeneous wire made of the core material. Analogously, it can easily be verified that the dispersion relations and the shell displacement (also the shell strain, not shown) match those of a bare wire made of the shell material in the limit $\gamma \to \infty$.

\section{Conclusions}
\label{sec:ConclusionsOutlook}

In conclusion, we have derived a comprehensive list of algebraic expressions that describe the static strain and the low-energy phonons in core/shell NWs. We take all stress and strain tensor elements into account, allow for arbitrary core and shell radii, and consider the elastic properties of the involved materials as independent. While the common approximation $\nu_c = \nu_s$ for Poisson's ratio in core and shell is often justified, we find that possibly important terms are ignored with this assumption [see, e.g., Eq.~(\ref{eq:erphicFlexCoreShell})].

We have investigated the resulting strain field for both the static and dynamical lattice displacement in great detail. Among other things, knowledge of the strain tensor elements is important for analyzing electron and hole spectra \cite{winkler:book, birpikus:book} and for studies that involve electron- and hole-phonon interactions \cite{stroscio:book, yu:book, roessler:book}. As seen in the example of Ge/Si NWs (Sec.~\ref{secsub:StatStrainResultsGeSiNWs}), the shell-induced strain can affect the carrier spectrum substantially \cite{kloeffel:prb11}. Furthermore, we have shown that the presence of a shell leads to additional, phonon-based strain components within the core that are absent in homogeneous NWs. Although the elements of the stress tensor are not listed in this work explicitly, they can directly be obtained via the stress-strain relations in Eqs.~(\ref{eq:StressStrainRelDiamondZB}) and (\ref{eq:StressStrainRelNW}). 

Given pseudomorphic growth, the dominant source of error in our model is certainly the assumption of isotropic materials ($c_{11} = c_{12} + 2 c_{44}$). Taking anisotropies exactly into account, however, is usually not possible without extensive numerical simulations \cite{hocevar:apl13, groenqvist:jap09, hestroffer:nanotech10} and, moreover, leads only to quantitative rather than qualitative corrections in most applications (see also Sec.~\ref{secsub:StatStrainAnalyticalResults}). We are therefore convinced that the results of our work will be very useful for future studies based on core/shell NWs.

\acknowledgments{
We thank P.\ Stano, S.\ Hoffman, and A.\ A.\ Zyuzin for helpful discussions and acknowledge support from the Swiss NF, NCCRs Nanoscience and QSIT, SiSPIN, DARPA, IARPA (MQCO), S$^3$NANO, and the NSF under Grant No.\ DMR-0840965 (M.T.).
}


\appendix

\section{Parameters for Ge/Si core/shell nanowires}
\label{sec:ParametersGeSiCoreShell}

In this appendix, we summarize the most important parameters used in the calculations for Ge/Si core/shell nanowires (NWs). The lattice constants of the core ($c$, Ge) and shell ($s$, Si) material are $a_c = 5.66\mbox{ \AA}$ and $a_s = 5.43\mbox{ \AA}$ \cite{winkler:book, adachi:properties}. The densities are $\rho_c = 5.32\mbox{ g/cm$^3$}$ and $\rho_s = 2.33\mbox{ g/cm$^3$}$ \cite{cleland:book, adachi:properties}. As explained below, the Lam\'{e} parameters are approximately $\lambda_c = 39.8\times 10^9\mbox{ N/m$^2$}$, $\mu_c = 55.6\times 10^9\mbox{ N/m$^2$}$, $\lambda_s = 54.5\times 10^9\mbox{ N/m$^2$}$, and $\mu_s = 67.5\times 10^9\mbox{ N/m$^2$}$.   

The elastic stiffness coefficients, taken from Ref.~\onlinecite{cleland:book} and listed here in units of $10^9\mbox{ N/m$^2$}$, are $c_{11}^c = 129$, $c_{12}^c = 48$, $c_{44}^c = 67.1$, $c_{11}^s = 165$, $c_{12}^s = 64$, and $c_{44}^s = 79.2$. We note that the numbers agree very well with those provided, e.g., in Ref.~\onlinecite{adachi:properties}. Introducing $p \in \{c, s\}$ for convenience and following Ref.~\onlinecite{kornich:prb14}, we approximate these stiffness coefficients by $\tilde{c}_{11}^{p}$, $\tilde{c}_{12}^{p}$, and $\tilde{c}_{44}^{p}$, respectively, such that the conditions $\tilde{c}_{11}^{p} = \tilde{c}_{12}^{p} + 2\tilde{c}_{44}^{p}$ for isotropic media are fulfilled. When the relative deviations of the three coefficients for a material are chosen to be the same, this approximation yields a relative deviation from the original values of 17.1\% in Ge and 14.8\% in Si. The results are $\tilde{c}_{12}^c = 39.8$, $\tilde{c}_{44}^c = 55.6$, $\tilde{c}_{12}^s = 54.5$, and $\tilde{c}_{44}^s = 67.5$, leading to the above-mentioned Lam\'{e} parameters due to $\tilde{c}_{12}^{p} = \lambda_p$ and $\tilde{c}_{44}^{p} = \mu_p$.

\section{Coordinate systems for stress and strain}
\label{sec:CoordSystemsStressStrain}

\subsection{Cartesian and cylindrical coordinates}
\label{secsub:CartandCylCoordSystems}

We mainly consider two different coordinate systems in this work. In the case of cartesian coordinates ($x$, $y$, $z$), a position $\bm{r}$ is described by
\begin{equation}
\bm{r} = x \bm{e}_x + y \bm{e}_y + z \bm{e}_z .
\end{equation}
The $z$ axis coincides with the symmetry axis of the NW, and $\bm{e}_x$, $\bm{e}_y$, and $\bm{e}_z = \bm{e}_x \times \bm{e}_y$ are the orthonormal basis vectors pointing along the axes indicated by subscripts. In cylindrical coordinates ($r$, $\phi$, $z$), we write
\begin{equation}
\bm{r} =  r \bm{e}_r + z \bm{e}_z ,
\end{equation}
with 
\begin{eqnarray}
x &=& r \cos{\phi} , \\
y &=& r \sin{\phi} ,
\end{eqnarray}
and so the $z$ axis is the same for both coordinate systems. We use
\begin{equation}
r = \sqrt{x^2 + y^2}
\end{equation}
throughout this work in order to avoid confusion with the density $\rho$. Therefore, $r \neq |\bm{r}|$. The unit vectors
\begin{eqnarray}
\bm{e}_r &=& \bm{e}_x \cos{\phi} + \bm{e}_y \sin{\phi} , \label{eq:erDefinitionAppendix}\\
\bm{e}_\phi &=& - \bm{e}_x \sin{\phi} + \bm{e}_y \cos{\phi} 
\label{eq:ephiDefinitionAppendix}
\end{eqnarray}
point in the radial and azimuthal direction, respectively, and we note that $\{\bm{e}_r, \bm{e}_\phi, \bm{e}_z\}$ forms again a right-handed, orthonormal basis.

\subsection{Transformation of second rank tensors}
\label{secsub:Transformation2ndRankTensors}

In the following, we recall how the strain and stress tensors transform under a change of basis and summarize the relations between the tensor elements in cartesian and cylindrical coordinates. The chosen notation is similar to the one in Ref.~\onlinecite{cleland:book}.
   
We consider two coordinate systems $\Sigma$ and $\Sigma^\prime$ with basis vectors $\bm{e}_i$ and $\bm{e}_i^\prime$, respectively, which are related via 
\begin{equation}
\bm{e}_j^\prime = \sum_i R_{ij} \bm{e}_i.
\label{eq:RotmatrixElementsRelationBasVecs}
\end{equation}
Given the two sets of basis vectors, an arbitrary vector $\bm{a}$ can be written as
\begin{equation}
\bm{a} = \sum_i a_i \bm{e}_i = \sum_i a_i^\prime \bm{e}_i^\prime ,
\label{eq:arbitraryVectora}
\end{equation} 
where $a_i$ ($a_i^\prime$) are the coefficients in $\Sigma$ ($\Sigma^\prime$). Inserting Eq.~(\ref{eq:RotmatrixElementsRelationBasVecs}) into Eq.~(\ref{eq:arbitraryVectora}) yields
\begin{equation}
a_j = \sum_i R_{ji} a_i^\prime .
\end{equation}
These linear relations between the coefficients of $\bm{a}$ can conveniently be written as (considering three dimensions)
\begin{equation}
\begin{pmatrix} a_1 \\ a_2 \\ a_3 \end{pmatrix} = R \begin{pmatrix} a_1^\prime \\ a_2^\prime \\ a_3^\prime \end{pmatrix} ,
\end{equation}
where the matrix
\begin{equation}
R = \begin{pmatrix}
R_{11} & R_{12} & R_{13} \\
R_{21} & R_{22} & R_{23} \\
R_{31} & R_{32} & R_{33} 
\end{pmatrix} 
\label{eq:RotationMatrixR3x3}
\end{equation}
comprises the elements $R_{ij}$ introduced in Eq.~(\ref{eq:RotmatrixElementsRelationBasVecs}). Since the basis vectors of $\Sigma$ and $\Sigma^\prime$, respectively, are orthonormal, one finds $R^{\rm T} R = \openone = R R^{\rm T}$, and so the transposed matrix $R^{\rm T}$ corresponds to the inverse operation, i.e., $R^{\rm T} = R^{-1}$. Consequently, the inverse relations between the coefficients $a_i$ and $a_i^\prime$ read
\begin{equation}
\begin{pmatrix} a_1^\prime \\ a_2^\prime \\ a_3^\prime \end{pmatrix} = R^{\rm T} \begin{pmatrix} a_1 \\ a_2 \\ a_3 \end{pmatrix} ,
\end{equation}
which is equivalent to
\begin{equation}
a_j^\prime = \sum_i R_{ij} a_i .
\end{equation}

The transformation rules for second rank tensors, such as stress and strain, can easily be found with the above-mentioned equations, keeping in mind that the result of a tensor acting on some vector $\bm{a}$ must be independent of the coordinate system. When the matrices $T^\prime$ and $T$, respectively, contain the tensor elements $T_{ij}^\prime$ and $T_{ij}$ for the primed and unprimed basis, the relations between these tensor elements are given in matrix form by
\begin{equation}
T^\prime = R^{\rm T} T R ,  
\label{eq:TprimeisRtTR}
\end{equation}
with $R$ as shown in Eq.~(\ref{eq:RotationMatrixR3x3}). An intuitive proof for this result is obtained by applying the left- and right-hand side to a column vector with coefficients $a_i^\prime$. As this column vector corresponds exactly to the representation of $\bm{a}$ in $\Sigma^\prime$, and analogously for the matrix $T^\prime$ representing the tensor, the left-hand side yields a column vector whose coefficients $b_i^\prime$ describe the result of the operation ``tensor on $\bm{a}$'' in the primed basis. On the right-hand side, the matrix $R$ transfers the coefficients $a_i^\prime$ of the column vector into $a_i$. Next, applying $T$ yields the resulting coefficients of ``tensor on $\bm{a}$'' in the unprimed basis. Finally, $R^{\rm T}$ replaces these coefficients $b_i$ by those in $\Sigma^\prime$, i.e., by $b_i^\prime$. The equality of left- and right-hand side holds for arbitrary $a_i^\prime$, and so Eq.~(\ref{eq:TprimeisRtTR}) applies. Analogously, 
\begin{equation}
T = R T^\prime R^{\rm T} .  
\label{eq:TisRTprimeRt}
\end{equation}            
We note that the results are equivalent to
\begin{eqnarray}
T_{m n}^\prime &=& \sum_{ij} R_{im} R_{jn} T_{ij} , \\
T_{m n} &=& \sum_{ij} R_{mi} R_{nj} T_{ij}^\prime .
\end{eqnarray}

Considering the cartesian and cylindrical coordinate systems introduced in Appendix~\ref{secsub:CartandCylCoordSystems} as $\Sigma$ and $\Sigma^\prime$, respectively, the matrix $R$ is
\begin{equation}
R = \begin{pmatrix}
\cos\phi & -\sin\phi  & 0 \\
\sin\phi & \cos\phi & 0 \\
0 & 0 & 1 
\end{pmatrix} .
\label{eq:RotationMatrixCartCyl}
\end{equation} 
Thus, exploiting Eq.~(\ref{eq:TprimeisRtTR}), the relations between the strain tensor elements in the two coordinate systems are 
\begin{gather}
\epsilon_{rr} = \epsilon_{xx} \cos^2\phi + \epsilon_{yy} \sin^2\phi + \epsilon_{xy} \sin(2 \phi) , \\
\epsilon_{\phi\phi} = \epsilon_{xx} \sin^2\phi + \epsilon_{yy} \cos^2\phi - \epsilon_{xy} \sin(2 \phi) , \\
\epsilon_{r\phi} = \epsilon_{xy} \cos(2\phi) + (\epsilon_{yy} -\epsilon_{xx}) \sin\phi \cos\phi , \\
\epsilon_{rz} = \epsilon_{xz} \cos\phi + \epsilon_{yz} \sin\phi , \\
\epsilon_{\phi z} = \epsilon_{yz} \cos\phi - \epsilon_{xz} \sin\phi , 
\end{gather}   
where we made use of $\epsilon_{ij} = \epsilon_{ji}$ and the trigonometric identities $\cos(2\phi) = \cos^2\phi - \sin^2\phi$ and $\sin(2\phi) = 2 \sin\phi \cos\phi$. The element $\epsilon_{zz}$ is the same in both coordinate systems, and the inverse relations
\begin{gather}
\epsilon_{xx} = \epsilon_{rr} \cos^2\phi + \epsilon_{\phi\phi} \sin^2\phi - \epsilon_{r\phi} \sin(2 \phi) , \\
\epsilon_{yy} = \epsilon_{rr} \sin^2\phi + \epsilon_{\phi\phi} \cos^2\phi + \epsilon_{r\phi} \sin(2 \phi) , \\
\epsilon_{xy} = \epsilon_{r\phi} \cos(2\phi) + (\epsilon_{rr} -\epsilon_{\phi\phi}) \sin\phi \cos\phi , \\
\epsilon_{xz} = \epsilon_{rz} \cos\phi - \epsilon_{\phi z} \sin\phi , \\
\epsilon_{yz} = \epsilon_{\phi z} \cos\phi + \epsilon_{rz} \sin\phi  
\end{gather} 
can be derived from Eq.~(\ref{eq:TisRTprimeRt}). The above-listed equations apply analogously to the stress tensor elements $\sigma_{ij}$, of course.

\subsection{Strain in cylindrical coordinates}
\label{secsub:StrainInCylindricalCoords}

In cylindrical coordinates, the relations between the displacement field 
\begin{equation}
\bm{u} = u_r \bm{e}_r + u_\phi \bm{e}_\phi + u_z \bm{e}_z 
\end{equation}
and the strain tensor elements $\epsilon_{ij}$ are nontrivial, since the basis vectors $\bm{e}_r$ and $\bm{e}_\phi$ depend on the angle $\phi$. Based on Eqs.~(\ref{eq:erDefinitionAppendix}) and (\ref{eq:ephiDefinitionAppendix}), one finds
\begin{eqnarray}
\partial_\phi \bm{e}_r &=& \bm{e}_\phi , \\
\partial_\phi \bm{e}_\phi &=& - \bm{e}_r .
\end{eqnarray}
Consequently,
\begin{gather}
\partial_r \bm{u} = \bm{e}_r \partial_r u_r + \bm{e}_\phi \partial_r u_\phi + \bm{e}_z \partial_r u_z , \\
\frac{1}{r} \partial_\phi \bm{u} = \bm{e}_r \frac{\partial_\phi u_r - u_\phi}{r} + \bm{e}_\phi \frac{\partial_\phi u_\phi + u_r}{r} + \bm{e}_z \frac{\partial_\phi u_z}{r} , \\
\partial_z \bm{u} = \bm{e}_r \partial_z u_r + \bm{e}_\phi \partial_z u_\phi + \bm{e}_z \partial_z u_z .
\end{gather}
With 
\begin{equation}
\epsilon_{ij}=\frac{1}{2}\left(\frac{\partial u_i}{\partial x_j}+\frac{\partial u_j}{\partial x_i}\right)
\end{equation}
as introduced in Sec.~\ref{sec:LinearElasticityTheory} of the main text, this yields \cite{landau:elasticity}
\begin{gather}
\epsilon_{rr} = \partial_r u_r , \\
\epsilon_{\phi\phi} = \frac{\partial_\phi u_\phi + u_r}{r} , \\
\epsilon_{zz} = \partial_z u_z  
\end{gather}
for the diagonal elements, and
\begin{gather}
\epsilon_{r\phi} = \frac{1}{2}\left(\frac{\partial_\phi u_r - u_\phi}{r}  + \partial_r u_\phi  \right) , \\
\epsilon_{rz} = \frac{1}{2}\left(\partial_z u_r + \partial_r u_z \right) , \\
\epsilon_{\phi z} = \frac{1}{2}\left(\partial_z u_\phi + \frac{\partial_\phi u_z}{r} \right) 
\end{gather}
for the off-diagonal elements.

\section{Phonons in bulk}
\label{sec:PhononsBulk}

In this appendix, we recall the theoretical description of acoustic phonons in bulk material. The formulas and results form the basis for Appendix \ref{sec:NormalizationNWs}, where we derive the normalization condition for phonons in core/shell and core/multishell NWs.  

\subsection{Plane waves and classical lattice vibrations}
\label{secsub:ClassicalLatticeVibsBulk}

We start from the dynamical equation of motion in an isotropic material \cite{cleland:book, landau:elasticity, roessler:book},
\begin{equation}
\rho \ddot{\bm{u}} = (\lambda + \mu) \nabla (\nabla \cdot \bm{u}) + \mu \nabla^2 \bm{u} ,  
\label{eq:DynEqIsotropicSolid}
\end{equation}
which is equivalent to 
\begin{equation}
\rho \ddot{u}_i = \sum_j \frac{\partial \sigma_{ij}}{\partial x_j} .
\label{eq:DynEqMostGeneral}  
\end{equation}
The introduced $\nabla$ is the Nabla operator, and $\nabla^2$ is the Laplacian. In bulk, where $\rho$, $\lambda$, and $\mu$ are constants, the elementary solutions of Eq.~(\ref{eq:DynEqIsotropicSolid}) are longitudinal ($l$) and transverse ($t_1$, $t_2$) plane waves with wave vectors $\bm{q}$,
\begin{eqnarray}
\bm{u}_{\bm{q}l}^\pm &\propto& \bm{e}_{\bm{q}l} e^{i(\bm{q} \cdot \bm{r} \pm \omega_{\bm{q} l} \tau)} , \\
\bm{u}_{\bm{q}t_1}^\pm &\propto& \bm{e}_{\bm{q}t_1} e^{i(\bm{q} \cdot \bm{r} \pm \omega_{\bm{q} t_1} \tau)} , \\
\bm{u}_{\bm{q}t_2}^\pm &\propto& \bm{e}_{\bm{q}t_2} e^{i(\bm{q} \cdot \bm{r} \pm \omega_{\bm{q} t_2} \tau)} ,
\end{eqnarray}
where the unit vectors $\bm{e}_{\bm{q}l} = \bm{q}/q$, $\bm{e}_{\bm{q}t_1} \perp \bm{e}_{\bm{q}l}$, and $\bm{e}_{\bm{q}t_2} = \bm{e}_{\bm{q}l} \times \bm{e}_{\bm{q}t_1}$ provide the polarizations and $\tau$ is the time. The dispersion relations between $q = |\bm{q}|$ and the angular frequencies $\omega_{\bm{q} l}$ and $\omega_{\bm{q} t} = \omega_{\bm{q} t_1} = \omega_{\bm{q} t_2}$ are
\begin{gather}
\omega_{\bm{q} l} = \sqrt{\frac{\lambda + 2 \mu}{\rho}} q =  v_l q , \\
\omega_{\bm{q} t} = \sqrt{\frac{\mu}{\rho}} q =  v_t q , 
\end{gather} 
respectively, where $v_l$ ($v_t$) is the speed of sound for longitudinal (transverse) polarization. Assuming periodic boundary conditions and a rectangular sample of length $L_x$, width $L_y$, and height $L_z$, the $N = V/a^3$ allowed values of $\bm{q} = q_x \bm{e}_x + q_y \bm{e}_y + q_z \bm{e}_z$ within the first Brillouin zone are given by
\begin{gather}
q_\kappa \in \left\{ -\frac{\pi}{a}, -\frac{\pi}{a} + \frac{2 \pi}{L_\kappa} , \cdots , \frac{\pi}{a} - \frac{2 \pi}{L_\kappa} \right\}, 
\label{eq:FirstBZqvaluesEven} \\
q_\kappa \in \left\{ -\frac{\pi}{a} + \frac{\pi}{L_\kappa}, -\frac{\pi}{a} + \frac{3 \pi}{L_\kappa} , \cdots , \frac{\pi}{a} - \frac{\pi}{L_\kappa} \right\} 
\label{eq:FirstBZqvaluesOdd}
\end{gather}
for even or odd $L_\kappa / a$, respectively, where $\kappa \in \{x, y, z \}$, $V = L_x L_y L_z$ is the sample volume, and $a$ is the lattice constant of the material.

Classically, an arbitrary acoustic lattice vibration in bulk is described by the displacement function
\begin{equation}
\bm{u}(\bm{r},\tau) = \sum_{\bm{q},s} \bm{e}_{\bm{q}s}  \left( c_{\bm{q}s} e^{i(\bm{q} \cdot \bm{r} - \omega_{\bm{q} s} \tau)} + \mbox{c.c.} \right) ,
\label{eq:uBulkClassicalGeneral}
\end{equation}
which corresponds to the most general, real-valued solution to Eq.~(\ref{eq:DynEqIsotropicSolid}) that fulfills the boundary conditions. The abbreviation ``c.c.'' stands for the complex conjugate, and the summation runs over all wave vectors $\bm{q}$ within the first Brillouin zone and the three mode types $s \in \{l, t_1, t_2 \}$. The real and imaginary parts of the complex coefficients $c_{\bm{q}s}$ (units: length) can be chosen according to the initial conditions, and we note that the $6 V/a^3$ free parameters in Eq.~(\ref{eq:uBulkClassicalGeneral}) are sufficient to set the initial positions and velocities of all $N = V/a^3$ lattice sites in the sample. Equation~(\ref{eq:uBulkClassicalGeneral}) yields the time derivatives 
\begin{eqnarray}
\dot{\bm{u}}(\bm{r},\tau) &=& - i \sum_{\bm{q},s} \omega_{\bm{q} s} \bm{e}_{\bm{q}s} \left( c_{\bm{q}s} e^{i(\bm{q} \cdot \bm{r} - \omega_{\bm{q} s} \tau)} - \mbox{c.c.} \right) , \label{eq:udotBulkClassicalGeneral} \\
\ddot{\bm{u}}(\bm{r},\tau) &=& - \sum_{\bm{q},s} \omega_{\bm{q} s}^2 \bm{e}_{\bm{q}s} \left( c_{\bm{q}s} e^{i(\bm{q} \cdot \bm{r} - \omega_{\bm{q} s} \tau)} + \mbox{c.c.} \right) ,
\label{eq:udotdotBulkClassicalGeneral}
\end{eqnarray}
which serve as input functions for the Hamiltonian that we discuss next.

\subsection{Hamiltonian}
\label{secsub:PhononsBasicHamiltonian}

In the continuum limit, the total energy of a lattice vibration comprises the kinetic energy 
\begin{equation}
E_{\rm kin} = \frac{1}{2} \int_V d^3\bm{r} \rho \dot{u}^2 
\end{equation}
and the elastic energy
\begin{equation}
U = \frac{1}{2} \sum_{i,j} \int_V d^3\bm{r} \sigma_{ij} \epsilon_{ij} ,
\label{eq:UtheGeneralOriginalIntegral}
\end{equation}
where we note that the density $\rho$ and the Lam\'{e} parameters $\lambda$ and $\mu$ depend on $\bm{r}$ in the general case. Integrating Eq.~(\ref{eq:UtheGeneralOriginalIntegral}) by parts and exploiting $\sigma_{ij} = \sigma_{ji}$ and Eq.~(\ref{eq:DynEqMostGeneral}), one finds that the elastic energy can be rewritten as
\begin{equation}
U = - \frac{1}{2} \int_V d^3\bm{r} \rho \bm{u} \cdot \ddot{\bm{u}} .
\end{equation}    
We omitted here the surface terms that arise from the integration by parts as these terms vanish in most cases due to the boundary conditions. For instance, considering the previously introduced, rectangular sample centered at the origin, it can directly be seen that the surface terms, which are of type
\begin{equation}
\left[ \sigma_{ij} u_i \right]_{x_j = - L_j/2}^{x_j = L_j/2} ,
\end{equation}
are zero both for periodic boundary conditions and for force-free sample surfaces. In the above example of the rectangular sample, the axes \{1,2,3\} correspond to $\{x,y,z\}$. We also verified that the surface terms vanish for the system discussed in Appendix~\ref{sec:NormalizationNWs}, i.e., for a cylindrically symmetric NW with a force-free surface perpendicular to the radial direction and periodic boundary conditions at the longitudinal ends. The resulting Hamiltonian 
\begin{equation}
H = E_{\rm kin} + U
\end{equation}
in the continuum limit reads
\begin{equation}
H = \frac{1}{2} \int_V d^3\bm{r} \rho \left( \dot{\bm{u}} \cdot \dot{\bm{u}} - \bm{u} \cdot \ddot{\bm{u}} \right). 
\label{eq:HgeneralWithUsContinuous}
\end{equation}  
With 
\begin{equation}
m(\bm{R}) = \rho a^3 |_{\bm{r} = \bm{R}}
\end{equation}
as the mass of the lattice site at position $\bm{R}$ (comprising the masses of all atoms in the corresponding unit cell), the discretized version of $H$ is
\begin{equation}
H = \frac{1}{2} \sum_{\bm{r} = \bm{R}} m \left( \dot{\bm{u}} \cdot \dot{\bm{u}} - \bm{u} \cdot \ddot{\bm{u}} \right), 
\label{eq:HgeneralWithUsDiscrete}
\end{equation} 
where the sum runs over all lattice vectors $\bm{R}$ in the material, i.e., over all lattice sites. 

We note in passing that insertion of Eqs.~(\ref{eq:uBulkClassicalGeneral}) to (\ref{eq:udotdotBulkClassicalGeneral}) of Appendix \ref{secsub:ClassicalLatticeVibsBulk} into Eq.~(\ref{eq:HgeneralWithUsContinuous}) or (\ref{eq:HgeneralWithUsDiscrete}) yields
\begin{equation}
H = 2 \rho V \sum_{\bm{q},s} |c_{\bm{q}s}|^2 \omega_{\bm{q} s}^2 ,
\end{equation}
which corresponds to the total energy of a classical acoustic lattice vibration in bulk.

\subsection{Quantization}
\label{secsub:QuantizationBulkPhonons}

\subsubsection{Generalized coordinates and momenta}
\label{secsubsub:GeneralizedCoordinates}

Before quantizing the phonon field, we demonstrate that $u_j(\bm{R})$ and
\begin{equation}
p_j(\bm{R}) = m(\bm{R}) \dot{u}_j(\bm{R})
\label{eq:generalizedMomentumPjR}
\end{equation}
correspond to the generalized coordinates and momenta of the lattice vibrations investigated here. The subscript $j$ refers to the three spatial directions and, as introduced before, $\bm{R}$ can be any of the $N$ lattice vectors of the material. Using the results of Appendix~\ref{secsub:PhononsBasicHamiltonian} and the Hamiltonian of Eq.~(\ref{eq:HgeneralWithUsDiscrete}), one finds
\begin{eqnarray}
\mathcal{L} &=& \sum_{\bm{R},j} p_j(\bm{R})\dot{u}_j(\bm{R}) - H \nonumber \\ 
&=& \frac{1}{2} \sum_{\bm{R}} m \left( \dot{\bm{u}} \cdot \dot{\bm{u}} + \bm{u} \cdot \ddot{\bm{u}} \right) = E_{\rm kin} - U ,
\end{eqnarray}
and so $\mathcal{L}$ indeed is the Lagrangian of the system. We note that 
\begin{equation}
\frac{\partial \mathcal{L}}{\partial \dot{u}_j(\bm{R})} = p_j(\bm{R})
\end{equation}
can easily be verified, since terms of type $\bm{u} \cdot \ddot{\bm{u}}$ are independent of $\dot{u}_j$ [see also Eqs.~(\ref{eq:DynEqIsotropicSolid}) and (\ref{eq:DynEqMostGeneral})]. Furthermore, as outlined below, one can prove that Hamilton's equations of motion are fulfilled, leaving no doubt about $u_j(\bm{R})$ and $p_j(\bm{R})$ being generalized coordinates and momenta.  

Equations (\ref{eq:DynEqIsotropicSolid}) and (\ref{eq:DynEqMostGeneral}) illustrate that $\ddot{\bm{u}}$ can be obtained from the displacement field $\bm{u}$ via spatial derivatives. Due to the linear dependence, one may write \cite{roessler:book}
\begin{gather}
\sum_{\bm{R}} \bm{u}(\bm{R}) \cdot m(\bm{R}) \ddot{\bm{u}}(\bm{R}) = \bm{U} \cdot D \bm{U} \nonumber \\
= \sum_{\bm{R},j} u_j(\bm{R}) \sum_{\bm{R}^\prime,j^\prime} D_{\bm{R},j,\bm{R}^\prime,j^\prime} u_{j^\prime}(\bm{R}^\prime) ,
\label{eq:UDUbasicequation}
\end{gather} 
where $\bm{U}$ is a $3N$-dimensional vector that comprises all $u_j(\bm{R})$, and $D$ is a $3N\mbox{$\times$}3N$ matrix whose matrix elements $D_{\bm{R},j,\bm{R}^\prime,j^\prime}$ do not depend on $\bm{U}$. An important relation evident from Eq.~(\ref{eq:UDUbasicequation}) is
\begin{equation}
\sum_{\bm{R}^\prime,j^\prime} D_{\bm{R},j,\bm{R}^\prime,j^\prime} u_{j^\prime}(\bm{R}^\prime) = m(\bm{R}) \ddot{u}_j(\bm{R}) = \dot{p}_{j}(\bm{R}).
\label{eq:DelementsRelationToPdot}
\end{equation}
Furthermore, the matrix $D$ must be Hermitian since $\bm{U} \cdot D \bm{U}$ is part of the Hamiltonian, and so 
\begin{equation}
D_{\bm{R},j,\bm{R}^\prime,j^\prime} = D_{\bm{R}^\prime,j^\prime,\bm{R},j} ,
\label{eq:DandDTelements}
\end{equation}
i.e., $D = D^{\rm T}$. We note in passing that the term $\bm{U} \cdot D \bm{U}$ and the property $D = D^{\rm T}$ apply to the case of real-valued $\bm{U}$ and $D$ considered here. More generally, with the complex conjugate denoted by an asterisk ($*$) and the Hermitian conjugate denoted by a dagger ($\dagger$), they correspond to $\bm{U}^* \cdot D \bm{U}$ and $D = D^\dagger$, respectively, and so $D_{\bm{R},j,\bm{R}^\prime,j^\prime} = D_{\bm{R}^\prime,j^\prime,\bm{R},j}^*$. Inserting Eq.~(\ref{eq:UDUbasicequation}) into Eq.~(\ref{eq:HgeneralWithUsDiscrete}) yields the Hamiltonian
\begin{equation}
H = \frac{1}{2} \sum_{\bm{R}, j}  m(\bm{R}) \dot{u}_j^2(\bm{R}) - \frac{1}{2} \bm{U} \cdot D \bm{U}  , 
\label{eq:HForHamiltonsEOMAppendix}
\end{equation} 
from which one finds 
\begin{equation}
- \frac{\partial H}{\partial u_j(\bm{R})} = \dot{p}_j(\bm{R})
\end{equation}
by exploiting Eqs.~(\ref{eq:DelementsRelationToPdot}) and (\ref{eq:DandDTelements}). Similarly,
\begin{equation}
\frac{\partial H}{\partial p_j(\bm{R})} = \dot{u}_j(\bm{R})
\end{equation}
can easily be verified with Eqs.~(\ref{eq:generalizedMomentumPjR}) and (\ref{eq:HForHamiltonsEOMAppendix}).

\subsubsection{Operators for phonons in bulk}
\label{secsubsub:OperatorsPhononsInBulk}

Having identified $u_j(\bm{R})$ and $p_j(\bm{R})$ as the generalized coordinates and momenta, we now promote them to operators and refer, e.g., to Ref.~\onlinecite{ashcroftmermin:book} for further details about the information summarized in this appendix. For the quantum mechanical description of  $u_j(\bm{R})$ and $p_j(\bm{R})$, we introduce the bosonic ladder operators $a_{\bm{q}s}^\dagger$ and $a_{\bm{q}s}$, which obey the commutation relations
\begin{gather}
\left[a_{\bm{q}s} , a_{\bm{q}^\prime s^\prime}^\dagger \right] = \delta_{\bm{q}, \bm{q}^\prime} \delta_{s, s^\prime} , \\
\left[a_{\bm{q}s}^\dagger , a_{\bm{q}^\prime s^\prime}^\dagger \right] = 0 = \left[a_{\bm{q}s} , a_{\bm{q}^\prime s^\prime} \right] , 
\end{gather} 
with $\delta_{\bm{q},\bm{q}^\prime}$ and $\delta_{s, s^\prime}$ as Kronecker deltas and $[A,B] = AB - BA$. We anticipate at this stage that the operators $a_{\bm{q}s}^\dagger$ and $a_{\bm{q}s}$ generate and annihilate, respectively, a phonon of mode $s$ with wave vector $\bm{q}$. Based on Eqs.~(\ref{eq:uBulkClassicalGeneral}) and (\ref{eq:udotBulkClassicalGeneral}), a reasonable ansatz for the operators $u_j(\bm{R})$ and $p_j(\bm{R})$ is
\begin{gather}
\bm{u}(\bm{R}) = \sum_{\bm{q},s} \bm{e}_{\bm{q}s}  \left( c_{\bm{q}s} a_{\bm{q}s} e^{i\bm{q} \cdot \bm{R}} + \mbox{H.c.} \right) , 
\label{eq:uRoperatorBulk} \\
\dot{\bm{u}}(\bm{R}) = \frac{\bm{p}(\bm{R})}{m} =  - i \sum_{\bm{q},s} \omega_{\bm{q}s} \bm{e}_{\bm{q}s}  \left( c_{\bm{q}s} a_{\bm{q}s} e^{i\bm{q} \cdot \bm{R}} - \mbox{H.c.} \right) ,
\label{eq:pRoperatorBulk}
\end{gather}
where $\bm{u}(\bm{R}) = \sum_j \bm{e}_j u_j(\bm{R})$, and analogous for the vector operators $\bm{p}(\bm{R})$ and $\dot{\bm{u}}(\bm{R})$. The Hermitian conjugate is denoted by ``H.c.''. As we analyze here the case of bulk material, the mass $m(\bm{R}) = m$ is independent of the lattice site.  

The introduced operators $u_j(\bm{R})$ and $p_j(\bm{R})$ need to meet a list of criteria. Of course, they must be Hermitian, which obviously is fulfilled. Moreover, they need to obey the canonical commutation relations
\begin{gather}
\left[ u_j(\bm{R}) , p_{j^\prime}(\bm{R}^\prime) \right] = i \hbar \delta_{j, j^\prime} \delta_{\bm{R}, \bm{R}^\prime} , 
\label{eq:canonicalCRpart1} \\
\left[ u_j(\bm{R}) , u_{j^\prime}(\bm{R}^\prime) \right] = 0 = \left[ p_j(\bm{R}) , p_{j^\prime}(\bm{R}^\prime) \right]  .
\label{eq:canonicalCRpart2}
\end{gather}
Exploiting, among other things, the identity
\begin{equation}
\sum_{\bm{q}} e^{i \bm{q} \cdot (\bm{R} - \bm{R}^\prime)} = N \delta_{\bm{R}, \bm{R}^\prime}
\end{equation}
and the properties of $\bm{e}_{\bm{q}s}$, particularly
\begin{equation}
\sum_s (\bm{e}_{\bm{q}s})_j (\bm{e}_{\bm{q}s})_{j^\prime} =  \delta_{j, j^\prime} 
\end{equation}  
due to their orthonormality, one finds that all canonical commutation relations apply when
\begin{equation}
|c_{\bm{q}s}|^2 = \frac{\hbar}{2 m N \omega_{\bm{q}s}} .
\label{eq:NormalizationConditionBulk}
\end{equation}
We note that $m N = \rho a^3 N = \rho V$ is the mass of the sample. The above-mentioned expressions for $u_j(\bm{R})$ and $p_j(\bm{R})$, however, are not the only ones that meet the criteria discussed so far. For instance, it is obvious that the canonical commutation relations also apply when $u_j(\bm{R}) \to u_j(\bm{R}) F_j(\bm{R})$ and $p_j(\bm{R}) \to p_j(\bm{R})/ F_j(\bm{R})$, where $F_j(\bm{R})$ is an arbitrary factor that is chosen real in order to preserve Hermiticity. Thus, an important criterion remains. We can ensure that the introduced operators feature the desired physical interpretations by verifying that the Hamiltonian takes the well-known form
\begin{equation}
H = \sum_{\bm{q},s} \hbar \omega_{\bm{q}s} \left( a_{\bm{q}s}^\dagger a_{\bm{q}s} + \frac{1}{2} \right) .
\label{eq:HamiltonianPhononsBulkQM}
\end{equation}  
Indeed, this requirement is satisfied for Eqs.~(\ref{eq:uRoperatorBulk}), (\ref{eq:pRoperatorBulk}), and (\ref{eq:NormalizationConditionBulk}). It may easily be proven by inserting these equations into Eq.~(\ref{eq:HgeneralWithUsDiscrete}), exploiting 
\begin{equation}
\ddot{\bm{u}}(\bm{R}) = - \sum_{\bm{q},s} \omega_{\bm{q}s}^2 \bm{e}_{\bm{q}s}  \left( c_{\bm{q}s} a_{\bm{q}s} e^{i\bm{q} \cdot \bm{R}} + \mbox{H.c.} \right) , 
\label{eq:udotdotRoperatorBulk}
\end{equation} 
the properties of $\bm{e}_{\bm{q}s}$ and $\omega_{\bm{q}s}$, and the identity
\begin{equation}
\sum_{\bm{R}} e^{i (\bm{q} - \bm{q}^\prime) \cdot \bm{R}} =  N \delta_{\bm{q}, \bm{q}^\prime} .
\end{equation} 
Equation~(\ref{eq:udotdotRoperatorBulk}) can readily be obtained from Eq.~(\ref{eq:uRoperatorBulk}) because the latter is written in terms of the eigenmodes, i.e., the eigenstates of the matrix $D$.   
We conclude that Eqs.~(\ref{eq:uRoperatorBulk}) and (\ref{eq:pRoperatorBulk}) are suitable operators for the canonical coordinates and momenta of acoustic lattice vibrations in bulk. The normalization condition is shown in Eq.~(\ref{eq:NormalizationConditionBulk}). 

Finally, we mention that the time dependence
\begin{equation}
O \to O(\tau) = e^{i H \tau / \hbar} O e^{- i H \tau / \hbar} 
\end{equation}
of an operator $O$, where $\tau$ is the time and $H$ is the phonon Hamiltonian [Eq.~(\ref{eq:HamiltonianPhononsBulkQM})], can easily be obtained via 
\begin{eqnarray}
a_{\bm{q}s}^\dagger &\to& a_{\bm{q}s}^\dagger(\tau) = a_{\bm{q}s}^\dagger e^{i \omega_{\bm{q}s} \tau}, \\
a_{\bm{q}s} &\to& a_{\bm{q}s}(\tau) = a_{\bm{q}s} e^{- i \omega_{\bm{q}s} \tau} .
\end{eqnarray}

\section{Phonon quantization in core/shell and core/multishell wires}
\label{sec:NormalizationNWs}

With the information of Appendix~\ref{sec:PhononsBulk}, the normalization condition for low-energetic phonons in cylindrically symmetric core/shell and core/multishell NWs can quickly be derived. Based on the main text, we consider here the four gapless phonon modes and denote the time-independent and unnormalized part of the corresponding displacement fields by 
\begin{equation}
\bm{u}_{q_z s}(\bm{r}) = \bm{u}_{q_z s}(r,\phi) e^{i q_z z} .
\label{eq:UOfVectorrToUOfrphiApp}
\end{equation} 
In contrast to Appendix~\ref{sec:PhononsBulk}, $s \in \{l, t, f_+, f_-\}$ now refers to the longitudinal ($l$) and torsional ($t$) modes, which fulfill
\begin{gather}
\bm{u}_{q_z l}(r,\phi) = \bm{u}_{q_z l}(r) = [\bm{u}_{q_z l}(r)]_r \bm{e}_r + [\bm{u}_{q_z l}(r)]_z \bm{e}_z , 
\label{eq:NWmodePropertiesLAppend} \\
\bm{u}_{q_z t}(r,\phi) = \bm{u}_{q_z t}(r) = [\bm{u}_{q_z t}(r)]_\phi \bm{e}_\phi, 
\label{eq:NWmodePropertiesTAppend} 
\end{gather} 
and to the two flexural modes ($f_\pm$), for which
\begin{equation}
\bm{u}_{q_z f_\pm}(r,\phi) = \bm{u}_{q_z f_\pm}(r) e^{\pm i \phi} .
\label{eq:NWmodePropertiesFAppend} 
\end{equation} 
The notation $\bm{u}_{q_z s}(r)$ indicates here that the components along $\bm{e}_r$, $\bm{e}_\phi$, and $\bm{e}_z$ depend solely on $r$ in cylindrical coordinates. In the main text, we explicitly show that the above-mentioned properties apply to homogeneous NWs and core/shell NWs, and we note that these features may analogously be considered for arbitrary core/multishell wires with cylindrical symmetry. The length of the NW is denoted by $L = L_z$, and $N_z = L_z/a$ is the number of lattice sites along $z$. We mention in passing that the lattice constant $a$ along the NW corresponds to an effective, equilibrated lattice spacing when the system is coherently strained (see, e.g., $a_e$ in Sec.~\ref{sec:StaticStrainMain} of the main text). Assuming periodic boundary conditions along $z$, the $N_z$ values of the wave number $q_z$ in the first Brillouin zone are given by Eqs.~(\ref{eq:FirstBZqvaluesEven}) and (\ref{eq:FirstBZqvaluesOdd}). Consequently, the displacement field of an arbitrary, classical lattice vibration based on the gapless modes reads
\begin{equation}
\bm{u}(\bm{r}, \tau) = \sum_{q_z , s} \left( c_{q_z s} \bm{u}_{q_z s}(\bm{r}) e^{- i \omega_{q_z s} \tau} + \mbox{c.c.} \right) ,
\label{eq:uNWClassical}
\end{equation}
where $\omega_{q_z s}$ (defined as positive) are the angular frequencies of the phonon modes, $c_{q_z s}$ are complex coefficients (dimensionless), and the sum runs over all $s$ and all $q_z$ within the first Brillouin zone. The time derivative is
\begin{equation}
\dot{\bm{u}}(\bm{r}, \tau) = - i \sum_{q_z , s} \omega_{q_z s} \left( c_{q_z s} \bm{u}_{q_z s}(\bm{r}) e^{- i \omega_{q_z s} \tau} - \mbox{c.c.} \right) .
\label{eq:udotNWClassical}
\end{equation}

The Hamiltonian derived in Appendix~\ref{secsub:PhononsBasicHamiltonian} and the discussion of the generalized coordinates and momenta in Appendix~\ref{secsubsub:GeneralizedCoordinates} are independent of the geometry and composition of the sample. (We note in passing that additional terms of type $\sigma_{ij}^{\rm static} \epsilon_{ij}^{\rm phonons}$ and $\sigma_{ij}^{\rm phonons} \epsilon_{ij}^{\rm static}$ arise for the elastic energy in the Hamiltonian of NWs when both the static (see also Sec.~\ref{sec:StaticStrainMain}) and the dynamical displacement are taken into account. However, these terms vanish when averaging over time and also when integrating over the NW. Contributions of type $\sigma_{ij}^{\rm static} \epsilon_{ij}^{\rm static}$ only provide a constant energy shift, and so we can focus here on the terms caused solely by phonons.) For the quantum mechanical description of low-energetic phonons in NWs, we therefore proceed analogously to Appendix~\ref{secsubsub:OperatorsPhononsInBulk} and define the operators
\begin{gather}
\bm{u}(\bm{R}) = \sum_{q_z,s} \left( c_{q_z s} a_{q_z s} \bm{u}_{q_z s}(\bm{R})  + \mbox{H.c.} \right) , 
\label{eq:uRoperatorNW} \\
\dot{\bm{u}}(\bm{R}) = \frac{\bm{p}(\bm{R})}{m(\bm{R})} =  - i \sum_{q_z,s} \omega_{q_z s}  \left( c_{q_z s} a_{q_z s} \bm{u}_{q_z s}(\bm{R}) - \mbox{H.c.} \right) .
\label{eq:pRoperatorNW}
\end{gather}
The creation ($a_{q_z s}^\dagger$) and annihilation ($a_{q_z s}$) operators for the respective phonons obey again the commutation relations
\begin{gather}
\left[a_{q_z s} , a_{q_z^\prime s^\prime}^\dagger \right] = \delta_{q_z, q_z^\prime} \delta_{s, s^\prime} , \\
\left[a_{q_z s}^\dagger , a_{q_z^\prime s^\prime}^\dagger \right] = 0 = \left[a_{q_z s} , a_{q_z^\prime s^\prime} \right] . 
\end{gather} 
As we focus only on the four gapless modes, the canonical commutation relations shown in Eqs.~(\ref{eq:canonicalCRpart1}) and (\ref{eq:canonicalCRpart2}) cannot be verified with Eqs.~(\ref{eq:uRoperatorNW}) and (\ref{eq:pRoperatorNW}) only. The reason is simply that the $c_{q_z s}$ introduced in Eq.~(\ref{eq:uNWClassical}) provide only $8 N_z$ free parameters, and so phonon modes of higher energy need to be included in order to reach the required $6 N$ degrees of freedom ($N$ is the number of lattice sites). Nevertheless, Eqs.~(\ref{eq:uRoperatorNW}) and (\ref{eq:pRoperatorNW}) can be considered consistent with the canonical commutation relations, given the formal analogy with the discussed case of bulk. The normalization condition can therefore be derived by choosing $c_{q_z s}$ such that the resulting Hamiltonian reads
\begin{equation}
H = \sum_{q_z , s} \hbar \omega_{q_z s} \left( a_{q_z s}^\dagger a_{q_z s} + \frac{1}{2} \right) .
\label{eq:HamiltonianPhononsNWQM}
\end{equation}

We start from the Hamiltonian in the continuum limit, Eq.~(\ref{eq:HgeneralWithUsContinuous}), and write
\begin{equation}
H = \frac{a}{2} \sum_z \int_0^{R_{\rm tot}} dr r \rho(r) \int_0^{2\pi} d\phi \Bigl[ \dot{\bm{u}}(\bm{r}) \cdot \dot{\bm{u}}(\bm{r}) - \bm{u}(\bm{r}) \cdot \ddot{\bm{u}}(\bm{r}) \Bigr] ,
\label{eq:HamStartForNormCondPhononsNW}
\end{equation}
where $\rho(\bm{r}) = \rho(r)$ because of the symmetry, and $R_{\rm tot}$ is the radius of the outermost shell, i.e., the total radius of the NW. The summation over the coordinate $z$ runs over the $N_z$ lattice positions along the $z$ axis. That is, we consider here a discrete lattice along the NW and use the continuum limit for the transverse directions. After insertion of [see also Eqs.~(\ref{eq:uRoperatorNW}) and (\ref{eq:pRoperatorNW})]
\begin{eqnarray}
\bm{u}(\bm{r}) &=& \sum_{q_z,s} \left( c_{q_z s} a_{q_z s} \bm{u}_{q_z s}(\bm{r})  + \mbox{H.c.} \right) ,  \\
\dot{\bm{u}}(\bm{r}) &=& - i \sum_{q_z,s} \omega_{q_z s}  \left( c_{q_z s} a_{q_z s} \bm{u}_{q_z s}(\bm{r}) - \mbox{H.c.} \right) , \\
\ddot{\bm{u}}(\bm{r}) &=& - \sum_{q_z,s} \omega_{q_z s}^2 \left( c_{q_z s} a_{q_z s} \bm{u}_{q_z s}(\bm{r})  + \mbox{H.c.} \right)  ,
\end{eqnarray}
and Eq.~(\ref{eq:UOfVectorrToUOfrphiApp}) into Eq.~(\ref{eq:HamStartForNormCondPhononsNW}), the dependence on $z$ can be removed due to \cite{ashcroftmermin:book}
\begin{equation}
\sum_z e^{i(q_z - q_z^\prime)z} = N_z \delta_{q_z, q_z^\prime} .
\end{equation} 
It is then convenient to introduce the shorthand notation
\begin{equation}
\mbox{Int}[\bm{a}\cdot \bm{b}] =  \int_0^{R_{\rm tot}} dr r \rho(r) \int_0^{2\pi} d\phi \hspace{0.03cm} \bm{a}\cdot \bm{b}  
\label{eq:IntDefinitionAppendix}
\end{equation}
with arbitrary $\bm{a}$ and $\bm{b}$, and we mention that $\mbox{Int}[\bm{a}\cdot \bm{b}] = \mbox{Int}[\bm{b}\cdot \bm{a}]$. Exploiting the properties of the phonon modes, in particular $\omega_{q_z s} = \omega_{-q_z s}$, $\omega_{q_z f_+} = \omega_{q_z f_-}$, and 
\begin{equation}
\mbox{Int}[\bm{u}_{q_z s}^{*}(r,\phi) \cdot \bm{u}_{q_z s^\prime}(r,\phi)] = \delta_{s, s^\prime} \mbox{Int}[\bm{u}_{q_z s}^{*} \cdot \bm{u}_{q_z s}]
\end{equation} 
due to Eqs.~(\ref{eq:NWmodePropertiesLAppend}) to (\ref{eq:NWmodePropertiesFAppend}), the Hamiltonian can be reduced to
\begin{equation}
H = 2 L_z \sum_{q_z , s} \omega_{q_z s}^2 |c_{q_z s}|^2 \mbox{Int}[\bm{u}_{q_z s}^{*} \cdot \bm{u}_{q_z s}] \left( a_{q_z s}^\dagger a_{q_z s} + \frac{1}{2} \right) .
\end{equation}
Thus, comparison with Eq.~(\ref{eq:HamiltonianPhononsNWQM}) yields the normalization condition
\begin{equation}
|c_{q_z s}|^2 \mbox{Int}[\bm{u}_{q_z s}^{*} \cdot \bm{u}_{q_z s}] = \frac{\hbar}{2 L_z \omega_{q_z s}} .
\end{equation}
For convenience, the arguments of $\bm{u}_{q_z s}$ were omitted here because of 
\begin{gather}
\bm{u}_{q_z s}^{*}(r,\phi) \cdot \bm{u}_{q_z s}(r,\phi) = \bm{u}_{q_z s}^{*}(r) \cdot \bm{u}_{q_z s}(r)   \nonumber \\
= \bm{u}_{q_z s}^{*}(\bm{r}) \cdot \bm{u}_{q_z s}(\bm{r}) = \bm{u}_{q_z s}^{*} \cdot \bm{u}_{q_z s},
\end{gather}
as evident from Eqs.~(\ref{eq:UOfVectorrToUOfrphiApp}) to (\ref{eq:NWmodePropertiesFAppend}). Insertion into Eq.~(\ref{eq:IntDefinitionAppendix}) therefore yields
\begin{equation}
\mbox{Int}[\bm{u}_{q_z s}^{*} \cdot \bm{u}_{q_z s}] = 2 \pi \int_0^{R_{\rm tot}} dr r \rho(r) \hspace{0.03cm} \bm{u}_{q_z s}^{*} \cdot \bm{u}_{q_z s} .
\end{equation}

In conclusion, the normalization condition for gapless phonon modes in cylindrically symmetric core/shell and core/multishell NWs is
\begin{equation}
|c_{q_z s}|^2 \int_0^{R_{\rm tot}} dr r \rho(r) \hspace{0.03cm} \bm{u}_{q_z s}^{*} \cdot \bm{u}_{q_z s} = \frac{\hbar}{4 \pi L_z \omega_{q_z s}} .
\label{eq:NormCondNWsFinalResAppendix}
\end{equation}
When $\rho(r) = \rho$, this condition matches the well-known result for homogeneous NWs \cite{stroscio:jap94}. In order to avoid confusion, we mention that the displacement operator $\bm{u}(\bm{r})$ is sometimes defined with an additional prefactor of $1/\sqrt{N}$. A general discussion of phonon quantization in nanostructures can be found in Ref.~\onlinecite{stroscio:book}. Analogous to the case of bulk (Appendix~\ref{secsubsub:OperatorsPhononsInBulk}), time-dependent operators may easily be obtained through $a_{q_z s}^\dagger \to a_{q_z s}^\dagger(\tau) = a_{q_z s}^\dagger e^{i \omega_{q_z s} \tau}$ and $a_{q_z s} \to a_{q_z s}(\tau) = a_{q_z s} e^{- i \omega_{q_z s} \tau}$.


\end{document}